\documentclass[aps,prd,nofootinbib,superscriptaddress]{revtex4} 
\usepackage{graphics}
\usepackage{epsfig}
\usepackage[english]{babel}
\usepackage{amsmath} 
\usepackage{verbatim} 
\usepackage{mathrsfs}
\usepackage{amsfonts}
\usepackage{amssymb}
\usepackage[latin1]{inputenc}
\usepackage{hyperref}

\newcommand{\tc}{\eta_{\vec{k}}^c}
\newcommand{\beq}{\begin{equation}}
\newcommand{\eeq}{\end{equation}}
\newcommand{\nk}{\textbf{k}}
\newcommand{\dphi}{\delta \phi}
\newcommand{\x}{\vec{x}}

\newcommand{\bra}{\langle}
\newcommand{\ket}{\rangle}

\newcommand{\mH}{\mathcal{H}}
\newcommand{\mP}{\mathcal{P}}

\newcommand{\barr}{\begin{eqnarray}}
\newcommand{\earr}{\end{eqnarray}}
\newcommand{\bea}{\begin{eqnarray*}}
\newcommand{\eea}{\end{eqnarray*}}

\newcommand{\nn}{\nonumber \\}

\begin{document}

\title{Primordial gravitational waves and the collapse of the wave function}

\author{Gabriel Le\'{o}n}
\affiliation{Departamento de F\'{\i}sica, Facultad de Ciencias Exactas y 
Naturales, Universidad de Buenos Aires, Ciudad Universitaria - PabI, Buenos 
Aires 1428, Argentina}  

\author{Lucila Kraiselburd}
\affiliation{Grupo de Astrof\'{\i}sica, Relatividad y Cosmolog\'{\i}a, Facultad 
de Ciencias Astron\'{o}micas y Geof\'{\i}sicas, Universidad Nacional de La 
Plata, Paseo del Bosque S/N 1900 La Plata, Pcia de Buenos Aires, Argentina}

\author{Susana J. Landau}
\affiliation{Departamento de F\'{\i}sica, Facultad de Ciencias Exactas y 
Naturales, Universidad de Buenos Aires and IFIBA, CONICET, Ciudad Universitaria 
- PabI, Buenos Aires 1428, Argentina\thanks{Member of  the Carrera del 
Investigador Cient\'{\i}fico y Tecnol\'ogico, CONICET}}

\begin{abstract}
 ``The self-induced collapse hypothesis'' was introduced by D. Sudarsky and 
collaborators to explain the origin of cosmic structure from a perfect isotropic 
and homogeneous universe during the inflationary regime. In this paper,  we 
calculate the power spectrum for the tensor modes, within the semiclassical 
gravity approximation, with the additional hypothesis of a generic self-induced 
collapse of the inflaton's wave function; we also compute an estimate for the 
tensor-to-scalar ratio. Based on this calculation, we show that the considered 
proposal exhibits a strong suppression of the tensor modes amplitude; 
nevertheless, the corresponding amplitude is still consistent with the joint 
BICEP/KECK and Planck Collaboration's limit on the tensor-to-scalar ratio. 
\end{abstract}


\maketitle

\section{Introduction}
\label{intro}

Observations of the Cosmic Microwave Background (CMB) radiation are one of the 
most powerful tools to study the early universe. Before the formation of neutral 
hydrogen, photons and electrons were coupled by Thomson scattering. Once 
photons were  decoupled from matter, they traveled freely through the universe 
almost without interacting with matter. Accordingly, the CMB radiation provides  
information about the physical parameters of the universe at recombination and 
also about the content of matter and energy density inside it. In the past 
two decades there has been a major improvement in the measurement of CMB 
fluctuations: a large number of experiments have been performed including a mix 
of of ground-, balloon- and space-based receivers.

Furthermore, from the estimation of the spectrum of initial perturbations, the 
CMB spectrum also provides an indirect evidence of the early universe physics. 
Different inflationary models proposed in the literature make distinct 
predictions of the primordial spectrum \cite{Martin14,infmodels} and, thus, the 
CMB radiation is an excellent observational tool to test them.  Temperature 
fluctuations are the result of perturbations in the gravitational potentials, 
which contribute directly to the fluctuations via gravitational red-shifting  
and 
which also drive acoustic oscillations of the primordial plasma. These 
processes 
result in temperature fluctuations that are of the same order of magnitude as 
the 
metric perturbations. On the contrary, CMB polarization is not directly 
generated by metric perturbations: a net polarization arises from Compton 
scattering only when the incident radiation field possesses a non-zero 
quadrupole moment. Polarization is only originated very near to the 
last scattering surface as the photons begin to decouple from the electrons 
generating a quadrupole moment through free-streaming. The tensor nature of CMB 
polarization allows a separation of scalar fluctuations from tensor ones.  CMB 
polarization maps can be decomposed into two terms: often-called $E$ and 
$B$-modes 
in analogy with the electric and magnetic field.  Gravitational waves generated 
during inflation, in contrast to adiabatic scalar perturbations, imprint a 
unique divergence-free pattern of polarization on the sky, namely, the 
$B$-mode.  
Scalar perturbations have no handedness so the $B$-mode component at low 
angular 
multipole scales  exists only if there is a tensor perturbation generated by 
primordial gravitational waves. On the other hand, detection of $B$-modes at 
large 
angular multipoles has been reported by the Polarbear experiment 
\cite{polarbear14}. However, it is known, that the source for this detection is 
gravitational lensing of the CMB radiation. In such a way, the 
 detection of the $B$-mode signal at low angular multipoles provides the 
cleanest 
window into the unique predictions of the inflationary cosmological paradigm. 
 
Last year, the BICEP collaboration reported a measurement of the $B$-mode 
polarization consistent with the prediction of standard inflationary models with 
tensor modes \cite{BICEP2}. However, it was pointed out  that, without an 
accurate dust map,  it is not possible to discern  between  dust polarization 
and polarization due to primordial gravity waves 
\cite{Mortonson14,Flauger14,Liu14}. In turn, an estimation of the dust present 
in BICEP2 experiment made by the Planck collaboration (extrapolating information 
from the 353 GHz map) showed that the detection informed by BICEP2 may be due to 
dust polarization \cite{planckdust}. Finally, a joint analysis of the BICEP/Keck 
and Planck collaborations \cite{PlanckBicep15} showed that the detection made by 
the BICEP collaboration is consistent with dust polarization and that there is 
no  statistically significant evidence for tensor modes.

According to the standard inflationary scenario, the onset of all cosmic 
structures  is explained  by considering a featureless stage described by a 
background Friedmann-Robertson-Walker (FRW) cosmology with a nearly exponential 
expansion driven by the potential of a single scalar field.\footnote{In the 
simplest models of inflation: $\phi$, the inflaton.} Additionally, the field's 
quantum fluctuations are characterized by a simple vacuum state, which is 
homogeneous and isotropic (for a proof of this statement see Appendix A of Ref. 
\cite{LLS13}).  In particular, the quantum fluctuations transmute into the 
classical statistical fluctuations that represent the seeds of the current 
cosmic structure like galaxies and galaxy clusters. However, there is an issue 
regarding the usual explanation for the origin of cosmic structure; this is,  
the standard inflationary paradigm lacks a physical mechanism capable of  
generating the inhomogeneity and anisotropy of our universe,  from an exactly 
homogeneous and isotropic initial state characterizing both: the background 
space-time and the quantum state of the inflaton. This problem has been 
analyzed 
in previous works  \cite{PSS06,Shortcomings,LLS13} and one key aspect of the 
problem is that there is no satisfactory solution within the standard physical 
paradigms of quantum unitary evolution because this kind of  dynamics is not 
capable to  break the initial symmetries of the system. To handle this 
shortcoming, a proposal has been developed by D. Sudarsky and collaborators 
\cite{PSS06,Sudarsky07,US08,Leon10,Leon11,DT11,LSS12,CPS13,LLS13,LLP14,Bispectrum}. In 
this 
scheme, a new ingredient is introduced into the inflationary scenario: \emph{the 
self-induced collapse hypothesis}. The main assumption is that,  at a certain 
stage in the cosmological evolution, there is an induced jump from the original 
quantum  state characterizing the particular mode of the quantum field; after 
the jump, the quantum state is inhomogeneous and anisotropic, namely,  it must 
not be an eigen-state of the linear and angular momentum operators. Therefore, 
by considering a self-induced collapse (in each mode) of the inflaton wave 
function, the inhomogeneities and anisotropies emerge at each particular length 
scale, i.e. the quantum collapse acts a source for the primordial curvature 
perturbation. As a consequence of this modification to the inflationary 
scenario, the predicted primordial power spectrum is changed as well as the CMB 
fluctuation spectrum.  Previous works \cite{PSS06,Shortcomings,LLS13,DT11} have 
extensively discussed both the conceptual and formal aspects of this  new 
proposal, and we refer the reader to the references. In this paper, motivated by the discussion generated around 
BICEP2's claim, we compute the  tensor power spectrum and the 
tensor-to-scalar ratio corresponding to the amplitude of primordial 
gravitational waves resulting from considering a generic self-induced collapse. 
Furthermore, in the theoretical framework of the self-induced collapse, the 
inflationary era is more accurately described by 
the semiclassical Einstein equations, namely, the perturbations of the scalar 
field are quantized and the metric perturbations remain always classical. In 
this paper, we calculate the  prediction for tensor primordial power spectrum  
generated by gravitational waves and find that the self-induced collapse 
hypothesis, plus semiclassical gravity,  predicts a strong suppression of the 
amplitude of the tensor modes.

It is worthwhile to mention that our result is consistent 
with the findings presented in Ref. \cite{Markkanen2014}, i.e. that the 
semiclassical gravity approximation plus a collapse of the inflaton's wave 
function lead those authors to qualitatively argue that the amplitude 
of the primordial gravitational waves
is undetectable; however, those authors consider a possible non-minimally 
coupling between gravity and the inflaton,
and also that the state collapses on a spacelike hypersurface for all 
wavelengths modes, this contrasts with our view
in which the time of collapse depends on the mode's wavelength. Also, in the 
present manuscript, we provide a detailed calculation of the tensor-to-scalar 
ratio (in Ref. \cite{Markkanen2014} the calculations involved only scalar 
perturbations).

The paper is organized as follows: In Section \ref{semicl}, we review the 
semiclassical gravity approximation and its relation with the self-induced 
collapse; in Section \ref{scalar}, we briefly review the proposal of the 
collapse of the wave function in the case where only scalar perturbations were 
considered; in Section \ref{tensor}, we present how the scalar perturbations, 
generated by the self-induced collapse, originates the tensor modes;  in Section 
\ref{spectrum}, we compute the power spectrum for the tensor modes; in Section 
\ref{estimation} we provide an estimate of the tensor-to-scalar ratio, and finally in 
Section \ref{conclusions}, we present our conclusions. We have also included an Appendix 
section in which we present the explicit form of some functions used in the calculations.

\section{The semiclassical gravity approximation and the collapse of the wave 
function}\label{semicl}

Our analysis of  the inflationary universe considers the corresponding  regime 
in a way  that a  quantum treatment of the matter fields is appropriate, while  
a classical description of gravitation would be justified simply because the  
measures of curvature are all well below the Planck scale. That is the realm of 
semiclassical gravity characterized  by  ($c=1$):

\beq
G_{ab} = 8 \pi G \bra \hat{T}_{ab} \ket. 
\eeq

The left hand side of this equation, namely the Einstein's tensor, contains the 
gravitational degrees of freedom that are always treated in a classical way. The 
right hand side, is the expectation value of a quantum operator, $\hat{T}_{ab}$ 
describing the quantum matter degrees of freedom.

Let us consider first how a self-induced collapse of the wave function fits 
into the semiclassical gravity framework (extended discussion can be found in 
Refs. \cite{DT11a,DT11}). It is known, that gravity is perhaps the most 
complicated  
 component of our universe to fit with the general paradigms offered by a 
quantum theory. There exists an extensive literature on this subject and the 
present manuscript does not even attempt to describe all the technical 
and/or conceptual problems associated with it. However, one aspect to consider 
is that, according to general relativity, the structure of the space-time is 
the core of gravitation, meanwhile, quantum theory seems to fit most easily in 
contexts where the space-time is already present (or given). 
In other words, quantum states are associated with objects that ``live'' in 
space-times. Evidently, deep conceptual modifications are in order if one 
aspires to provide a characterization of space-time itself in a quantum 
language. For instance, the canonical approach to quantum gravity  
\cite{DeWitt67,Kieffer06}, suffers from the so called problem of time 
\cite{Isham92}. It seems thus natural to 
speculate that in this setting a dramatic departure from the quantum orthodoxy, 
such as a dynamical reduction of the wave function might find its origin, 
i.e. it is possible that a more fundamental description of quantum gravity 
would take the form of deviations form the standard unitary evolution that 
characterize quantum theory as we know it (for a particular example how these 
ideas can be implemented in the black hole information paradox see Refs. 
\cite{Elias,Sujoy,Sujoy2}). 

In other words, it seems a plausible conjecture that a variation of standard 
quantum theory, that we have referred to here as described effectively by the 
``collapse of the wave function,'' corresponds to lasting features of the 
fundamental timeless (and probably spaceless) theory of quantum gravity. If 
that is the case, the emergence of space-time itself would be tied to the 
incorporation of such effective quantum description of matter fields living on a
space-time, and evolving approximately according to standard quantum field 
theory on curved spaces, with some small deviations that might include our 
collapse hypothesis. Similar ideas of this kind regarding gravity as an 
emergent phenomenon have been considered previously (e.g. Refs. 
\cite{jacobson,ashtekar}). This seems to suggest that, in the context where we 
consider the collapse of the wave function, the space-time itself must be 
considered as an approximate phenomenological characterization, and, therefore, 
something that cannot be subjected to quantization. For a more pictorial 
example, we can consider the following analogy: the propagation of heat in a 
medium. This phenomenon can be described by the heat equation $\partial T/ 
\partial t - \nabla^2 T = S$, where $T$ is the temperature of the medium and $S$ 
the heat sources. It is quite clear that despite the fact that this equation 
resembles an equation for some field, it would be meaningless to quantize it. 
Furthermore, we can consider some situation in which the source of heat requires 
a quantum mechanical treatment, i.e. taking $S$ as a quantum operator. Under 
such conditions, it seems natural that to the extent that the temperature 
description is still relevant and of interest, the right hand side of the 
equation above should be replaced by something like $\bra \hat S \ket$. 
Evidently, there will be situations that are so far removed from the context 
where the heat equation was derived that even the notion of temperature itself 
would become meaningless. We equally expect that in the fundamental quantum 
theory of gravity we will be able to find several situations where the 
semiclassical gravity approximation would fail completely, but in following 
with our line of thought and previous analogy, it seems quite possible that 
those would correspond to situations where the concept of space-time itself 
becomes meaningless. 

Clearly, all these arguments above are filled with ``educated guesses'' and 
conjectures, and we only take them as guidances. Nevertheless, as we will show 
in the rest of this manuscript, when considering these ideas within the context 
of the inflationary universe and, in particular, using them to analyze the 
emergence of primordial gravitational waves, we are able to provide an 
observational constraint. If it happens that our proposal does not match the 
observational data, then clearly we should find another way to incorporate 
our self-induced collapse proposal in the traditional inflationary setting, 
namely, the one involving the quantization of metric perturbations (some 
preliminary steps in that direction can be found in Refs. \cite{MBL14,LB15}). 

After the previous digression, we turn our focus on the generation of the 
primordial perturbations provided by the self-induced collapse.

The   starting point provided 
by inflation corresponds to  an homogeneous and isotropic state for the 
gravitational and matter degrees of freedom; this is, the background space-time 
is symmetric and the quantum state characterizing the inflaton is an 
eigen-state 
of the linear and angular momentum operators. The self-induced collapse 
proposal, states that at some stage during inflation, the quantum state of the 
matter fields, normally associated to the inflaton, undergoes  a spontaneous  
collapse (evidently without relying on any external entities such as 
``observers,'' ``measurement devices,'' etc.). The resulting  state of the 
matter fields no longer needs to  share the symmetries of the initial state, 
and, its coupling  to  the  gravitational degrees of freedom, through  
Einstein's semiclassical equation, leads to a geometry that is no longer 
homogeneous and isotropic. Thus, the quantum collapse acts as a source for the 
inhomogeneities and anisotropies in the universe, this is the main role for 
introducing the collapse of the wave function. 

The idea of a self-induced collapse of the wavefunction, in a non-cosmological 
setting, has been analyzed in great detail in the past (for a review of 
objective collapse models see Ref.  \cite{bassi2003}). Furthermore, R. Penrose 
and L. Di\'{o}si have long advocated to gravity as the main agent triggering 
the 
collapse \cite{penrose1996,diosi1987}; the reasoning is that a self-induced 
collapse occurs when the matter fields are in a quantum superposition that 
would 
lead to corresponding space-time geometries that are ``too distinct among 
themselves.'' This kind of gravity-induced collapse would be happening in 
fairly 
common situations, providing a resolution of the measurement problem in quantum 
mechanics.  In our view, which can be thought as an attempt to apply these 
ideas 
to the cosmological context, in order to address the shortcoming mentioned in 
the Introduction, the present formalism must be regarded as an effective 
description of the fundamental collapse mechanism, possibly as consequence of a 
successful quantum gravity theory, which in the present situation leads to a 
transition from the symmetric vacuum state to the non-symmetrical latter state 
(the symmetry being homogeneity and isotropy). We will not discuss further 
these 
motivations here and instead our analysis should be regarded as a purely 
phenomenological scheme, in the sense that it does not seek to provide a 
physical mechanism for the collapse, but merely to present a generic 
parametrization of the collapse. The main formalism and conceptual framework 
can 
be consulted in Refs. \cite{PSS06,Shortcomings,LLS13,DT11}.

There is also an issue regarding the semiclassical approximation during the 
collapse. It is well known that introducing a dynamical collapse generically 
violates the conservation of energy, so the divergence of the energy-momentum  
tensor $\nabla_a \bra \hat{T}^{ab} \ket$ does not vanish. If that happens, then 
of course the divergence of the Einstein tensor does not vanish either, which 
evidently is a problem since we know that this divergence must be zero. 
Therefore, during the collapse, we cannot say how the modified dynamics, 
provided by a dynamical reduction of the wave function associated to the 
quantum fields, affects the classical 
metric perturbations that are directly related to the observables, that is, the 
temperature anisotropies. However, this shortcoming does not necessarily mean 
that we cannot implement a collapse mechanism in our formalism; the 
semiclassical gravity approximation is valid before  and after the collapse and 
these will be the two regimes of interest for the present work. Before 
 the collapse, the source for the metric perturbations is zero and only after 
the collapse (where there is no violation of energy conservation)  the 
expectation values of the operator $\hat{T}_{ab}$ evaluated at the 
post-collapse 
state, act as a source for the metric perturbations.

As we mentioned, the principal role for the collapse of the wave function is to 
act as the main agent for breaking the symmetries (homogeneity and isotropy) of 
the primordial universe. At first-order, the scalar metric perturbations are 
generated, via the semiclassical approximation, by the expectation value of the 
matter scalar perturbations in the post-collapse state.
Nevertheless, the matter perturbations--at this order--do not act as a source 
for the tensor metric perturbations, thus, within our model, the amplitude for 
first-order tensor modes is zero. In turn, second-order tensor modes, can be 
generated via first-order scalar perturbations (of both metric and matter 
perturbations) 
\cite{Mollerach1997,Mollerach1998,Mollerach2003,Osano2006,Ananda2006,Baumann2007}.  
As a consequence, one naively expects that the tensor to scalar ratio computed 
within 
the semiclassical gravity approximation  will be much smaller than the one 
computed in the standard case. On the other hand, as shown in Refs. 
\cite{PSS06,Sudarsky07,US08,Leon10,Leon11,DT11,LSS12,CPS13,LLS13}, the 
self-induced collapse modifies the scalar power spectrum in a very particular 
way, namely by including a function of the time of collapse,  and in principle, 
we would not know \emph{a priori} if the same modification to the tensor power 
spectrum might result in a detectable amplitude of primordial gravitational 
waves even if the tensor power 
spectra is generated at second-order in the perturbations. This is different 
from the standard procedure where the Mukhanov-Sasaki variable (which mixes 
metric and inflaton's perturbations) is quantized. In that case, tensor modes 
are generated at the first-order approximation; however, as analyzed in Refs. 
\cite{Osano2006,Ananda2006,Baumann2007}, one can also compute the second-order 
contribution to the tensor modes provided by scalar perturbations.

We would like to emphasize the fact that tensor perturbations are exactly 
zero at first-order because we are considering the semiclassical Einstein equations 
and not because the assumption of a self-induced collapse of the inflaton's wave 
function. In fact, in recent papers, one of us has calculated the tensor power 
spectrum  and the tensor-to-scalar ratio working in terms of a joint 
metric-matter quantization and including the self-induced collapse of the 
inflaton's wave function leading to different results as the ones reported here 
\cite{MBL14,LB15}. In particular, we will assume that the collapse of the wave 
function of each mode of the inflaton field has occurred creating first-order 
scalar metric perturbations; then, we will use such perturbations as a source 
of second-order tensor modes, i.e. as the origin of the primordial 
gravitational waves. In the following sections, we will estimate the amplitude 
of the second-order tensor modes generated by the process described
  above.

\section{Inflation and the collapse of the wave function: scalar perturbations}

\label{scalar}

In this section we present a very brief review of the implementation of the 
collapse proposal to the inflationary universe, in particular, we will focus on 
the first-order scalar perturbations of the metric. Since this subject has been 
extensively discussed in previous works 
\cite{PSS06,US08,Leon10,Shortcomings,DT11,LSS12,LLS13}, we will only present 
the 
main results, and refer to the reader to the aforementioned works for more 
details.

We  start with  the standard  inflationary model characterized  by   the action 
($c=1$): 
\begin{equation}\label{accioncolapso}
S[\phi,g_{ab}] = \int d^4x \sqrt{-g} \bigg( \frac{R[g] }{16 \pi G} - 
\frac{1}{2} 
\nabla_a \phi \nabla^a \phi- V[\phi] \bigg).
\end{equation}
with $\phi$ a scalar field representing the inflaton. The metric and scalar 
field  are separated into  background plus perturbations. The background is 
represented by a spatially flat FRW space-time with line element $ds^2 = 
a(\eta)[-d\eta^2 + \delta_{ij}dx^i dx^j]$ and the homogeneous part of the 
scalar 
field $\phi_0 (\eta)$ in the slow-roll regime.  The scale factor corresponding 
to the inflationary era is $a(\eta) \simeq -1/(H \eta)$ with $H$ the Hubble 
factor  which,  during inflation, is related to the inflaton potential as $H^2 
\simeq (8 \pi G /3) V \simeq$ constant. From these considerations, one infers 
that the conformal time $\eta$ is in the range $\eta \in (-\infty,\eta_r)$, 
where $\eta_r$ is very small in absolute terms, $\eta_r \simeq -10^{-22}$  Mpc. 
The slow-roll regime,  is characterized  by   $\phi_0' \simeq -(a^3/3a') 
\partial_\phi V$,  and the  slow-roll parameter  is $\epsilon \equiv 
1-\mH'/\mH^2 \simeq \frac{1}{2} M_P^2 (\partial_\phi V/V)^2  \ll 1$; here, 
 a prime denotes partial derivative with respect to conformal time $\eta$ and  
$M_P$ is the reduced Planck's mass defined as $M_P^2 \equiv 1/(8 \pi G)$ 
($\hbar=1$). Furthermore, we will work with the approximation $\epsilon =$ 
constant.

Let us focus on  the first-order scalar perturbations, as these will be the 
ones 
of interest for the present work. We will be working in the so-called 
longitudinal gauge,\footnote{The analysis  is done  by choosing a specific  
gauge and not  in terms of the so  called ``gauge invariant quantities,'' this 
is because in the picture followed  here, the  metric  and field fluctuations 
are  treated on a different footing. The inflaton's field perturbations, are 
given a standard quantum field (in curved space-time) treatment, while the 
metric 
is considered as a classical object that describes in an effective manner the 
deeper fundamental degrees of freedom of the quantum gravity theory that one is 
believed to lie underneath; the two descriptions are related through Einstein's 
semiclassical equations. The choice of  gauge  implies that the time  
coordinate 
is attached  to  some specific  slicing of  the  perturbed space-time, and 
thus, 
our identification of  the corresponding  hypersurfaces (those of constant 
time) 
 as the  ones  associated with the  occurrence of collapses,--something deemed 
as an actual physical  change--,  turns  what is normally  a simple  choice of  
gauge  into a choice of   the distinguished  hypersurfaces,  tied to the 
putative physical process behind  the collapse. This  naturally leads to  
tensions  with the expected   general covariance of a fundamental theory, a  
problem that afflicts  all known  collapse  models, and  which in the 
non-gravitational   settings becomes the issue   of compatibility with  Lorentz 
or Poincare invariance of the proposals.  We must acknowledge that this generic 
problem  of  collapse  models is indeed  an open issue for the present 
approach. One  would expect that its  resolution  would be tied to the  
uncovering the  actual  physics  behind  what we treat here as  the  collapse 
of the  wave function (we which we  view as a merely an effective 
description).} 
the perturbed metric is, thus, represented by:

\beq
ds^2=a(\eta)^2[-(1+2\Phi)d\eta^2+(1-2\Psi)\delta_{ij}dx^idx^j].
\eeq
In the absence of anisotropic stress $\Phi=\Psi$. The quantity $\Psi(\x,\eta)$ 
is called the ``Newtonian potential'' and it also represents the curvature 
perturbation in the longitudinal gauge.

Turning our attention to the inflaton's perturbations, it is convenient to   
work with the quantum field  $\hat{y}=a \hat{\dphi}$,  and its canonical 
conjugate momentum $\hat{\pi}=a \hat{\dphi}'=\hat{y}'-\hat{y}a'/a$. For each 
Fourier component of the perturbations,  Einstein's semiclassical equations at 
first-order $\delta G_{ab}^{(1)} = 8\pi G  \bra \delta \hat{T}_{ab}^{(1)} \ket$ 
yield
\begin{equation}\label{master2}
\Psi_{\nk} (\eta) = \sqrt{\frac{\epsilon}{2}} \frac{H}{M_P k^2} \bra 
\hat{\pi}_{\nk} (\eta) \ket.
\end{equation}
It is clear from Eq. \eqref{master2} that if the state of the field is the 
vacuum state, all modes associated to the metric perturbation vanish, and, thus 
the space-time is homogeneous and isotropic.

As  discussed in Ref. \cite{PSS06,US08,Leon11}, as  a result of  the  collapse,  
each mode  jumps  to  a new state $| \Theta \ket$ characterized  by  the 
expectation value of a certain operator, which is determined by the pre-collapse 
 uncertainties and a random number. In  the scheme  we  consider here, the state 
  will be  (partially)  characterized by: 
\begin{equation}
\langle \hat{y}_{\vec{k}}^{R,I} (\eta_k^c) \rangle_\Theta = 0,  \quad \langle 
\hat{\pi}_{\vec{k}}^{R,I} (\eta_k^c) \rangle_\Theta = x_{\vec{k}}^{R,I} 
\sqrt{[\Delta \hat{\pi}_{\vec{k}}^{R,I} (\tc)]_0^2}, 
\end{equation}
where $R,I$ denotes the real and imaginary parts of the operators, respectively; 
$\tc$ represents the \emph{time of collapse} for each mode. The pre-collapse  
state  is the  Bunch-Davies vacuum and   $[\Delta 
\hat{\pi}_{\vec{k}}^{R,I}]_0^2$  is the corresponding uncertainty.   This    
scheme is motivated  by  the fact  that the variable which is directly related 
with the Newtonian potential $\Psi$, is the expectation value of $\hat \pi$, but 
it also serves to simplify, without loss of generality, the second-order 
calculation of the next section.

The   $x_{\nk}^{R,I}$   are numbers  selected randomly  from  a  Gaussian 
distribution    centered  at $0$ and  with  unit dispersion..  In our approach,  
our universe  corresponds to a \textbf{single realization} of these random 
variables, and, thus,  each of these quantities  has a  single  specific value.  
Some  statistical   aspects  concerning these    quantities can  be studied   
using as  a tool an imaginary ensemble of ``possible universes,'' but we  should 
  in principle distinguish   those   from  the statistics of  such  quantities 
for the  particular  universe  we inhabit.

In terms of the random variables, Eq. \eqref{master2} can be rewritten as 
\cite{PSS06,US08,Leon11}
\begin{equation}\label{masterrandom}
\Psi_{\nk} (\eta) =     \frac{ \sqrt{\epsilon}  H}{ M_P} \left( \frac{L}{2k} 
\right)^{3/2}   \left( \cos [k\eta - z_k]+ \frac{\sin [k\eta - z_k]}{z_k} 
\right) X_{\nk}, 
\end{equation}
where $X_{\nk} \equiv x_{\nk}^R + i x_{\nk}^I$, $z_k \equiv k\tc$ and $\tc$ is 
the time of collapse of each mode $k$, also $L$ is the side length of the cubic 
box in which we are performing the quantization of the fields. In other words, 
the Fourier's modes $\Psi_{\nk}$ correspond to $\Psi (\eta,\vec x) = 1/L^3 
\sum_{\nk \neq 0} \Psi_{\nk} (\eta) e^{i \nk \cdot \x}$, where the sum is over 
wave vectors $\nk$ satisfying $k_i L = 2 \pi n_i$ for $i = 1,2,3$ with $n_i$ 
integer; at the end of calculations we can take the limit $L \to \infty$ thus 
making $\nk$ continuous.

This result expresses  the Newtonian potential,  after the collapse,  in terms 
of the random variables  that determine the post-collapse state. One of the  
advantages of our  approach  is that the  origin of the randomness, which one   
usually attributes to  quantum theory,   becomes   transparent and  specific:  
the variables  $X_{\nk}$  characterize, once  and  for all,  every kind of 
stochasticity involved. The  Newtonian potential is closely  related with the  
observational quantity  i.e., the  temperature anisotropies. Therefore,   the  
set of all the random variables $\lbrace X_{\nk_1}, X_{\nk_2}, X_{\nk_3} \ldots  
\rbrace$ corresponding to  our universe fixes   the  value  of the  observed 
temperature anisotropies.  Naturally,  we cannot give a definite prediction for 
those  values,  however,  as  we will show next,   the fact that   a large 
number  of modes  $\vec k$   contribute to  the observed  temperatures  
anisotropies,  justifies a statistical  analysis that leads
  to  a theoretical estimate for the observational  quantities.

Furthermore, $\Psi_{\nk}$ also depends on the time of collapse $\tc$ (recall the 
definition $z_k \equiv k\tc$), which is the additional parameter introduced by 
our model. 

In previous works \cite{US08,LSS12} it has been shown that if $z_k$ is 
independent of $k$ our model yields the same prediction for the scalar 
primordial power spectrum as the  standard inflationary models and, thus, the 
predictions for the CMB temperature, $E$-mode and $TE$ cross-correlation spectrum is 
also the same as in the standard scenario. In fact, in \cite{Pia2015}, it is shown that 
the prediction for the scalar power spectrum, which is obtained from Eq. 
\eqref{masterrandom}, is

\beq\label{PS}
P_S (k,\eta) \simeq \frac{H^2}{64 \pi^2 M_P^2 \epsilon} \left( \cos [z_k] - 
\frac{\sin[z_k]}{z_k}\right)^2 k^{n_s-1}.
\eeq

Moreover, in Ref. \cite{LSS12} some 
of us  have also performed a statistical analysis using WMAP 7-year-data 
together with other CMB data and constrained the possible departures of the assumption 
$z_k$ independent of $k$. On the other hand,  this shows that implementing 
the collapse hypothesis to the inflationary universe, translates into adding a 
new degree of freedom, namely the time of collapse $ \tc$.  Therefore, in what 
follows, we will assume that $z_k$ is a constant which implies $k \tc = z$ with 
$z$ a free parameter of the collapse model. In other words, the relation between $k$ and 
$\tc$ is just phenomenological, since at present there is no workable covariant 
formulation of the collapse mechanism that could give us a physical justification for 
such relation. In summary, from now on we will use the relation $\tc \propto k^{-1}$, 
which is equivalent to $z_k \equiv k \tc = z$ independent of $k$.

\section{Collapse induced gravitational waves}
\label{tensor}

Let us now turn to the case of the tensor perturbations. In this section, we will be 
following closely the mathematical results presented in Refs. 
\cite{Acquaviva2002,Osano2006,Ananda2006,Baumann2007} but bearing in mind that the 
physical situation in our case differs from the previous works.

The line element associated to 
the metric at first-order in the tensor perturbations,  is given by

\beq
ds^2 = a^2(\eta)[-d\eta^2  + (\delta_{ij}+h^{(1)}_{ij})dx^idx^j]
\eeq

Relying on Einstein's (first-order) perturbed equations,  $\delta 
G_{ij}^{(1)}  = 8\pi G  \bra \delta \hat{T}_{ij}^{(1)} \ket$, one can obtain the motion 
equation for the tensor perturbations

\beq\label{masterh1}
h^{(1)''}_{ij}+2\mH h^{(1)'}_{ij}-\nabla^2 h^{(1)}_{ij}=0.
\eeq

Comparing Eqs. \eqref{master2} and \eqref{masterh1}, we observe that the 
collapse of the wave function does not affect equally the first-order scalar and 
tensor perturbations. The expectation value (in a post-collapse state) of the 
momentum of the inflaton generates the first-order scalar perturbations, while, 
the first-order tensor perturbations does not contain a similar source (i.e. 
there are no terms such as 
$\partial_i \bra \hat \dphi \ket \partial_j \bra \hat 
\dphi' \ket$ ); hence, $h^{(1)}_{ij} =0$ even after the collapse has taken 
place.

Focusing now on the second-order metric perturbations  and working in the generalized 
longitudinal gauge \cite{Acquaviva2002},  the metric's components can be 
described by

\begin{subequations}
 \begin{equation}
  g_{00}= -a(\eta)^2 (1+2\Phi^{(1)} + \Phi^{(2)})
 \end{equation}
\begin{equation}
 g_{0i}=0
\end{equation}
\beq
g_{ij} = a(\eta)^2 \left[ (1-2\Psi^{(1)}-\Psi^{(2)}) \delta_{ij} + \frac{1}{2} 
(\partial_i V^{(2)}_j  + \partial_j V^{(2)}_i)  + \frac{1}{2} h^{(2)}_{ij}  \right]
\eeq
\end{subequations}
where $\Psi^{(1)}, \Phi^{(1)}$, $\Psi^{(2)}, \Phi^{(2)}$ correspond to the first 
and second-order scalar perturbations respectively, while  
$h^{(2)}_{ij}$ corresponds to the second-order tensor perturbations; we have also 
included the second-order vector perturbations $V_i^{(2)}$. 

As we mentioned previously, by assuming anisotropic stress one has $\Psi^{(1)} = 
\Phi^{(1)}$. Einstein's second-order perturbed equations, $\delta G^{i(2)}_j = 8\pi G  
\delta T^{i(2)}_j $ yields  \cite{Acquaviva2002,Ananda2006,Baumann2007}

\barr
& &\bigg[ \frac{1}{4}(h^{i(2)''}_j + 2\mH h^{i(2)'}_j - \partial_k \partial^k 
h^{i(2)}_j) + 4 \Psi^{(1)} \partial^i \partial_j \Psi^{(1)} + 2 \partial^i 
\Psi^{(1)} \partial_j \Psi^{(1)} + (\textrm{$\Psi^{(2)}, \Phi^{(2)}, V_i^{(2)}$ terms})   
\nonumber \\
&+&  ( \textrm{Diag. part including terms 
$\Psi^{(1)}$,$\Psi^{(2)}$,$\Phi^{(2)}$} ) \delta^i_j \bigg] \nonumber \\
&=& 4 \pi G \bigg[  \bigg( \delta \phi^{(2)'} \phi_0' - \delta \phi^{(2)} 
\frac{\partial V}{\partial \phi} a^2 + (\delta \phi^{(1)'})^2 - \partial_k 
\delta \phi^{(1)} \partial^k \delta \phi^{(1)}+ 4 (\Psi^{(2)})^2 \phi_0'^2 
\nonumber \\
&-& (\delta \phi^{(1)})^2  \frac{\partial^2 V}{\partial \phi^2} a^2 - 
4\Psi^{(1)} \dphi^{(1)'} \phi_0' \bigg) \delta^i_j + 2 \partial^i \dphi^{(1)} 
\partial_j \dphi^{(1)} \bigg].
\earr


We note that $h^{(2)}_{ij}$ corresponds to a symmetric, transverse and traceless 
tensor. Consequently, in order to obtain the equations for the second-order tensor 
perturbations, one can construct a projection tensor ${\mP_{ij}}^{lm}$ 
\cite{Ananda2006, Baumann2007} (to be defined explicitly next) that extracts the 
transverse, traceless part of any tensor. Applying the projection tensor 
${\mP_{ij}}^{lm}$ on both sides of Einstein's second-order perturbed equations eliminates 
the contribution from the diagonal terms and from the objects $\Psi^{(2)}, \Phi^{(2)}, 
V_i^{(2)}$. 

From now on we will change the notation slightly: We will omit the index 
$^{(1)}$ from first-order scalar perturbations, and since we are interested in 
second-order tensor perturbations, we will also omit the index $^{(2)}$ from 
$h_{ij}^{(2)}$.  

Consequently, the operation ${\mP_{ij}}^{lm} \delta G^{(2)}_{lm} = 
8\pi G  {\mP_{ij}}^{lm} \bra \hat \delta T^{(2)}_{lm} \ket$, yields

\beq\label{masterh2}
h_{ij}'' + 2\mH h_{ij}' - \nabla^2 h_{ij} = -4 \mP_{ij}^{lm} S_{lm},
\eeq
with $S_{ij}$ is defined as

\beq\label{Sij}
S_{ij} \equiv 4 \Psi \partial_i \partial_j \Psi + 2 \partial_i \Psi \partial_j 
\Psi - 8 \pi G \partial_i \bra \hat \dphi \ket \partial_j \bra \hat \dphi \ket. 
\eeq

Thus, unlike Eq. \eqref{masterh1}, the equation for second-order tensor 
perturbations, Eq. \eqref{masterh2}, contains a source term provided by 
$S_{ij}$. This is, after the collapse, $\Psi$ is no longer zero, and act as 
source for $h_{ij}$; also, the source term $S_{ij}$ contains the expectation 
value of the quantum degrees of freedom, i.e. the quantum inhomogeneities of the 
inflaton $\bra \hat \dphi \ket$.

We will proceed to solve Eq. \eqref{masterh2}; in order to show explicitly how 
the collapse proposal comes into play. Let us recall that Einstein's first-order 
perturbed equations with components $(0i)$ yield

\beq
\partial_ i ( \mH \Psi + \Psi') = 4 \pi G \phi_0' \partial_i \bra \hat \dphi 
\ket.
\eeq
Using Friedmann's equations and the slow-roll approximation, one shows that $4 
\pi G \phi_0' = - \sqrt{\epsilon/2} \mH /M_P$; accordingly, last equation yields:

\beq\label{0i}
\partial_i \bra \hat \dphi \ket= - \sqrt{\frac{2}{\epsilon}} \frac{M_P}{\mH} 
\left(\mH \partial_i \Psi + \partial_i \Psi'      \right).
\eeq
Then, by substituting Eq. \eqref{0i} into Eq. \eqref{Sij}, $S_{ij}$ is given by

\beq
S_{ij} = -2 \partial_i \Psi \partial_j \Psi \left( 1+ \frac{1}{\epsilon} \right) 
- \frac{4}{\epsilon \mH} \partial_i \Psi \partial_j \Psi'    - \frac{2}{\epsilon 
\mH^2} \partial_i \Psi' \partial_j \Psi'.
\eeq

We define the Fourier transform of the tensor metric perturbations as

\beq
h_{ij} (\x,\eta) = \int \frac{d^3 \nk}{(2\pi)^{3/2}} e^{i \nk \cdot \x} [h_{\nk} 
(\eta) e_{ij} (\nk) +  \bar h_{\nk} (\eta)  \bar e_{ij} (\nk)].  
\eeq
where we defined two time-independent polarization tensors $e_{ij}$ and $\bar 
e_{ij}$ that may be expressed in terms of orthonormal basis vectors $e_i$, $\bar 
e_j$ orthogonal to $\nk$, explicitly

\beq
e_{ij} (\nk) \equiv \frac{1}{\sqrt{2}} [ e_i (\nk) e_j (\nk) - \bar e_i (\nk) 
\bar e_j (\nk) ], 
\eeq

\beq
\bar e_{ij} (\nk) \equiv \frac{1}{\sqrt{2}} [ e_i (\nk) \bar e_j (\nk) + \bar 
e_i (\nk)  e_j (\nk) ].
\eeq

In terms of these polarization tensors, the projection tensor ${\mP_{ij}}^{lm}$ is 
defined as

\beq
{\mP_{ij}}^{lm} S_{lm} \equiv  \int \frac{d^3 \nk}{(2\pi)^{3/2}} e^{i \nk \cdot 
\x}  [ e_{ij} (\nk) e^{lm} (\nk) + \bar e_{ij} (\nk) \bar e^{lm} (\nk) ] S_{lm} 
(\nk) 
\eeq
where

\beq
S_{lm} (\nk) = \int \frac{d^3 \x'}{(2\pi)^{3/2}} e^{-i \nk \cdot \x'} S_{lm} 
(\x').
\eeq
Therefore, in Fourier space, the equation of motion for the amplitude of the 
tensor mode $h_{\nk}$ is

\beq\label{hkmaster1}
h_{\nk}'' + 2\mH h_{\nk}' +k^2 h_{\nk} = S(\nk,\eta),
\eeq
where

\beq
S(\nk,\eta) = 4 \int \frac{d^3 \tilde \nk }{(2\pi)^{3/2}} e^{lm} (\nk) \tilde 
k_l \tilde k_m \bigg[ 2 \bigg(1+ \frac{1}{\epsilon} \bigg) \Psi_{\tilde \nk} 
(\eta) \Psi_{\nk-\tilde \nk} (\eta) + \frac{4}{\epsilon \mH}   \Psi_{\tilde \nk} 
(\eta) \Psi_{\nk-\tilde \nk}' (\eta) + \frac{2}{\epsilon \mH^2} \Psi_{\tilde 
\nk}' (\eta) \Psi_{\nk-\tilde \nk}' (\eta)       \bigg].
\eeq

By performing a change of variables $v_{\nk} \equiv a h_{\nk}$, Eq. 
\eqref{hkmaster1} becomes

\beq\label{vkmaster}
v_{\nk}''+ \left( k^2 - \frac{a''}{a} \right) v_{\nk} = a S(\nk,\eta).
\eeq
In order to solve Eq. \eqref{vkmaster}, we can rely on the Green's function 
method, thus

\beq\label{hksol}
h_{\nk} (\eta) = \frac{1}{a(\eta)} \int_{-\infty}^{\infty} d\tilde \eta \:
 g_{k} (\eta;\tilde \eta) a(\tilde \eta) S(\nk, \tilde \eta).
\eeq
Green's function $g_{k} (\eta,\tilde \eta)$ satisfies  

\beq\label{eqgreen}
g_{k}'' + \left( k^2 - \frac{a''}{a}   \right)g_{k} = \delta (\eta-\tilde 
\eta),
\eeq
where $\eta>\tilde \eta$ and the primes indicate derivative with respect to $\eta$.

One can find exact solutions to Eq. \eqref{eqgreen} by using that during 
inflation $a \simeq -1/(H\eta)$, in this way, $a''/a \simeq 2/\eta^2$. Thereupon, the 
retarded Green's function is 

\beq\label{green}
g_{k} (\eta,\tilde \eta) = -k\eta \tilde \eta [ j_1 (k\eta) y_1 (k \tilde 
\eta) -  j_1 (k\tilde \eta) y_1 (k\eta)] \Theta(\eta-\tilde \eta),
\eeq
with $\Theta(\eta-\tilde \eta)$ the step function and  $j_1,y_1$ spherical Bessel's 
functions of first and second kind of order 1 respectively. 


Furthermore, by making use of the Green's solution and the fact 
that during inflation $\mH \simeq -1/\eta$, $1+1/\epsilon \simeq 1/\epsilon$, then 
Eq. \eqref{hksol} is

\beq\label{hksol2}
h_{\nk} (\eta) = -k\eta^2 \int_{\tc}^{\eta} d\tilde \eta 
S(\nk,\tilde \eta) [ j_1 (k\eta) y_1 (k \tilde \eta) -  j_1 (k\tilde \eta) y_1 
(k\eta)]
\eeq
 and  $S(\nk, \tilde \eta)$ is rewritten

\beq\label{sk}
S(\nk,\tilde \eta) = \frac{8}{\epsilon}  \int \frac{d^3 \tilde \nk 
}{(2\pi)^{3/2}} e^{lm} (\nk) \tilde k_l \tilde k_m [\Psi_{\tilde \nk} ( \tilde 
\eta) \Psi_{\nk-\tilde \nk} ( \tilde  \eta)-2\Psi_{\tilde \nk} ( \tilde \eta) 
\Psi_{\nk-\tilde \nk}' ( \tilde \eta) +  \tilde \eta^2 \Psi_{\tilde \nk}' ( 
\tilde  \eta) \Psi_{\nk-\tilde \nk}' ( \tilde  \eta)].
\eeq

Note that the range of integration in Eq. \eqref{hksol2} is from $\tc$ to $\eta$ 
(additionally, during inflation $-\infty < \eta < 0$). We remind the reader that before 
the time of collapse $\tc$ there are no perturbations of the metric at any scale and the 
space-time is exactly homogeneous and isotropic. That is, for $\eta<\tc$ the source 
term $S(k,\eta)$ is zero and, consequently, $h_{\nk} (\eta)$ is also zero. It is only 
after the self-induced collapse has taken place that the scalar perturbations of the 
metric are born (hence $S(k,\eta)\neq 0$) and, in turn, they act as a source for the 
(second-order) tensor perturbations. 

At this point, we insert our expression for $\Psi_{\nk}$ given by the collapse 
proposal, namely Eq. \eqref{masterrandom} [the derivative $\Psi_{\nk}'$ can be 
calculated also from Eq. \eqref{masterrandom}] into Eq. \eqref{sk}. Therefore, 
the amplitude of the gravitational wave $h_{\nk}$ is

\beq\label{hkmaster}
h_{\nk} (\eta)= -2k\eta^2 \frac{H^2}{M_P^2} \int \frac{d^3 \tilde \nk 
}{(2\pi)^{3/2}} e^{lm} (\nk) \tilde k_l \tilde k_m \frac{X_{\tilde \nk} X_{\vec 
q} L^3}{\tilde k^{3/2} q^{3/2} } \int_{\tc}^{\eta} d \tilde \eta 
G(k\eta,k\tilde \eta) F(\tilde k \tilde \eta, q \tilde \eta,z),
\eeq
where $\textbf{q} \equiv \nk - \tilde \nk$, thus, $ q = | \nk - \tilde \nk|$,

\beq
G(k\eta,k\tilde \eta) \equiv j_1 (k\eta) y_1 (k \tilde \eta) -  j_1 (k\tilde 
\eta) y_1 (k\eta),
\eeq

\barr
 F(\tilde k \tilde \eta, q \tilde \eta,z) &\equiv& \cos [q \tilde \eta] \cos 
[\tilde k \tilde \eta] (R_1(z)^2 + R_2(z)^2 \tilde k q \tilde \eta^2) + \sin [q 
\tilde \eta] \sin [\tilde k \tilde \eta] (R_1(z)^2 \tilde k q \tilde \eta^2+ 
R_2(z)^2 ) \nonumber \\
 &+& 2q \tilde \eta \{ R_1(z)^2  \cos [\tilde k \tilde \eta] \sin [q \tilde 
\eta]  -  R_2(z)^2 \sin [\tilde k \tilde \eta] \cos [q \tilde \eta] \} \nonumber 
\\
 &+&  R_1(z) R_2(z)\{- 2q \tilde \eta \cos[ (\tilde k + q) \tilde \eta] + ( 1-  
\tilde k q  \tilde \eta^2) \sin [ (\tilde k + q)\tilde \eta] \}, 
\earr
and 

\beq
R_1 (z) \equiv  \cos [z]- \frac{\sin [z]}{z}  \qquad  R_2 (z) \equiv \sin [z] + 
\frac{\cos [z]}{z}.
\eeq
We have also made an abuse of notation because $\Psi_{\nk}$, as expressed in Eq. 
\eqref{masterrandom}, was obtained by performing the quantization for the 
inflaton field  in a cubic box of side $L$ with discrete $\nk$; so, there is an 
$L^3$ in the expression for $h_{\nk}$. However, in the following sections, when 
we compute the observational quantities, we will take the limit $L \to \infty$, 
which assure us that $\nk$ becomes continuous. 

Equation \eqref{hkmaster} is the main result of this section. It explicitly 
relates the parameters characterizing the collapse, i.e. the random variables 
$X_{\nk}$ and the time of of collapse $z \equiv k \tc$ (through the functions 
$R_1 (z)$, $R_2 (z)$). We recall that $h_{\nk} (\eta)$ is related to the metric 
perturbation corresponding to second-order tensor modes; as argued previously, 
first-order tensor modes are zero within the semiclassical gravity 
approximation.

We note that the random variables $X_{\nk}$ are fixed once the collapse has 
occurred (or to be more precise, once a given collapse mechanism has ended). If 
we somehow knew how the collapse mechanism is related to each mode $\nk$, we 
could perform the integral over $\tilde \nk$, and give a definite prediction for 
$h_{\nk}$. However, even if we cannot give such definite prediction, we can use 
the statistical properties of the random variables $X_{\nk}$ to make contact 
with the observations, namely, the power spectrum for the gravitational waves, 
this will be the subject of the next section. Nevertheless, we can see clearly 
how the randomness of the metric perturbations is inherited by the stochasticity 
of the collapse of the wave function codified in the random variables 
$X_{\nk}$.

Focusing now on the function $F(\tilde k \tilde \eta, q \tilde \eta,z)$, we note 
that it includes the functions $R_1^2(z)$, $R_2^2(z)$ and $R_1(z) R_2 (z)$.
We remind the reader that if one assumes  $z_k 
\equiv k\tc$ independent of $k$ (i.e. $\tc \propto k^{-1}$), then  the prediction for the 
scalar power spectrum is the same as the standard one, see Eq. \eqref{PS}, which  fits 
very well the recent observational data. Hence, from now on, we will use $z_k=z$ 
independent of $k$.

\section{The power spectrum for gravitational waves within the collapse model}
\label{spectrum}

In this section, we will make contact with the observational quantities. This 
is, we will calculate the power spectrum for the gravitational waves $ P_T 
(k,\eta)$, defined as

\beq\label{defps}
\overline{h_{\nk} (\eta) h_{\nk'}^* (\eta)} \equiv \frac{2 \pi^2}{k^3} P_T 
(k,\eta) \delta (\nk-\nk') 
\eeq

The bar appearing in $\overline{h_{\nk} (\eta) h_{\nk'}^* (\eta)}$ denotes an 
average over possible realizations of $h_{\nk}$. In our approach,  $h_{\nk}$ 
depends directly on the random variables $X_{\nk}$, thus, in the collapse 
framework, a realization of $h_{\nk}$ is provided by the self-induced collapse, 
which in turn yields a single realization for $X_{\nk}$, the stochasticity is 
naturally inherited by the randomness of the collapse.  Moreover, the set of all 
modes associated to the random field $\{h_{\nk_1}, h_{\nk_2}, h_{\nk_3}, \ldots  
  \}$ characterizes a particular universe $\mathcal{U}$. Thus, the average is 
over possible realizations describing different universes $\mathcal{U}_1, 
\mathcal{U}_2, \ldots$ Our universe, is just one particular materialization 
$\mathcal{U}^*$. We want to remark that this is different form the standard 
inflationary account, in which the power spectrum is normally obtained from the 
Fourier's transform of the quantum two-point function $\bra 0 | \hat h_{\nk} 
(\eta) \hat h_{\nk'} (\eta) | 0 \ket$. Then, in the traditional approach, 
somehow (e.g. by invoking decoherence, squeezing of the vacuum, many-world 
interpretation of quantum mechanics, etc.) occurs the transition $\hat h_{\nk} 
\to h_{\nk} = A e^{i \alpha_{\nk}}$ with $\alpha_{\nk}$ a random phase and $A$ 
is identified with the quantum uncertainty of $\hat h_{\nk}$, i.e. $A^2=\bra 0 | 
\hat h_{\nk}^2 |0\ket$, but the random nature of $h_{\nk}$ remains unclear. In 
our approach, $h_{\nk}$ is always a classical quantity, before the collapse is 
zero, it is only after the collapse that $h_{\nk} \neq 0$, but the metric 
perturbation never undergoes a sort of quantum-to-classical transition that 
needs to be justified.

Therefore, by using $h_{\nk} (\eta)$ from Eq. \eqref{hkmaster}, we have

\barr\label{psh1}
\overline{h_{\nk} (\eta) h_{\nk'}^* (\eta)} &=& 4k'k\eta^4 \frac{H^4}{M_P^4}\int 
\frac{d^3 \tilde \nk }{(2\pi)^{3/2}} \int \frac{d^3 \tilde \nk' }{(2\pi)^{3/2}} 
e^{lm} (\nk) \tilde k_l \tilde k_m e^{rs} (\nk') \tilde k'_r \tilde k'_s 
\frac{1}{(\tilde{k} q\tilde{k'}q')^{3/2}} \nonumber \\
&\times& \int_{\tc}^{\eta} d\tilde \eta_1 
\int_{\tc}^{\eta} d\tilde \eta_2   G(k\eta,k\tilde \eta_1) 
G(k'\eta,k'\tilde \eta_2) F(\tilde k \tilde \eta_1, q \tilde \eta_1,z) F(\tilde 
k' \tilde \eta_2, q' \tilde \eta_2,z)    \nonumber \\
&\times & L^6 \overline{X_{\tilde \nk} X_{\textbf{q}} X_{\tilde \nk'}^* X_{\vec 
q'}^*}.
\earr

Since we have assumed that the variables $X_{\nk}$ are Gaussian distributed, 
then 

\beq
\overline{X_{\tilde \nk} X_{\textbf{q}} X_{\tilde \nk'}^* X_{\textbf{q}'}^*} =  
\overline{X_{\tilde \vec  k} X_{\textbf{q}}} \times \overline{X_{\tilde \vec  k'}^* 
X_{\textbf{q}'}^*} + \overline{X_{\tilde \vec k} X_{\tilde \vec k'}^*} \times 
\overline{X_{\textbf{q}} X_{\textbf{q}'}^*} + \overline{X_{\tilde \vec k } X_{\vec 
q'}^*} \times \overline{X_{\textbf{q}}. X_{\tilde \vec k'}^*}.
\eeq
Moreover, $\overline{X_{\nk} X_{\nk'}^*} = 2 \delta_{\nk,\nk'} $ and   
$\overline{X_{\nk} X_{\nk'}} = \overline{X_{\nk}^* X_{\nk'}^*} = 2 
\delta_{\nk,-\nk'} $. We note that $\delta_{\nk,\nk'}$ refers to  Kronecker's 
delta, which reflects the fact that we have performed the quantization of the 
inflaton in a cubic box with volume $L^3$. We proceed to take the limit $L \to 
\infty$ making $\nk$ continuous. In this limit the Kronecker's delta goes to a 
Dirac's delta, i.e. $L^3 \delta_{\nk,\nk'} \to \delta (\nk-\nk')$. Thus,

\beq\label{deltas}
L^6 \overline{X_{\tilde \nk} X_{\textbf{q}} X_{\tilde \nk'}^* X_{\textbf{q}'}^*} \to 2 
[\delta (\tilde \nk + \textbf{q}) \delta(\tilde \nk' + \textbf{q}') + \delta(\tilde \nk 
- \tilde \nk') \delta(\textbf{q} - \textbf{q}') + \delta (\tilde \nk - \textbf{q}') 
\delta 
(\textbf{q} - \tilde \nk')     ].
\eeq
 
Substituting Eq. \eqref{deltas} into Eq. \eqref{psh1} and integrating over 
$\tilde \nk '$ yields

\barr
\overline{h_{\nk} (\eta) h_{\nk'}^* (\eta)}  &=& 8 k^2 \eta^4 \frac{H^4}{M_P^4} 
\int \frac{d^3 \tilde \nk}{(2\pi)^3}  \frac{(e^{lm} (\nk) \tilde k_l \tilde 
k_m)^2}{(\tilde k q)^3} \int_{\tc}^{\eta} d\tilde \eta_1 
\int_{\tc}^{\eta} d\tilde \eta_2  G(k\eta,k\tilde \eta_1) 
G(k\eta,k\tilde \eta_2)  F(\tilde k \tilde \eta_1, q \tilde \eta_1,z) \nonumber 
\\
&\times& [F(\tilde k \tilde \eta_2 , q \tilde \eta_2,z) + F(q\tilde \eta_2 , 
\tilde k \tilde \eta_2,z)    ] \delta(\nk-\nk').
\earr
From this last expression, we can extract the power spectrum for the 
gravitational waves. After a change of variables,\footnote{The change of 
variables corresponds to: $x\equiv k\eta, z \equiv k \tc,\tilde x_1 
\equiv k \tilde \eta_1, \tilde x_2 \equiv  k\tilde \eta_2, v\equiv \tilde k/ k, 
u \equiv q/k$} the power spectrum is given by

\barr\label{psgw1}
P_T (k,\eta) &=&  x^4 \frac{H^4}{8\pi^4M_P^4} \int_{z}^{x} 
d\tilde x_1 d\tilde x_2 G(x,\tilde x_1) G(x,\tilde x_2) \nonumber \\
&\times& \int_0^\infty dv \int_{|v-1|}^{|v+1|} du \frac{[4v^2 - (u^2-v^2 
-1)^2]^2}{u^2v^2} F(v\tilde x_1, u \tilde x_1,z) [ F(v\tilde x_2, u \tilde 
x_2,z) + F(u\tilde x_2, v \tilde x_2,z)    ], 
\earr
where we have multiplied by a factor of 2 the power spectrum by taking into 
account the polarization of the gravitational wave and we have used that $e^{lm} 
(\nk) \tilde k_l \tilde k_m = \tilde k^2 ( 1 - \cos \theta)$, with $\theta$ the 
angle between $\nk$ and $\tilde \nk$. By taking into account the change of 
variables, the expressions for $G$ and $F$ are:

\beq
G(x,\tilde x) = j_1(x) y_1 (\tilde x) - j_1 (\tilde x) y_1 (x),
\eeq
 
and

\barr
F(v\tilde x, u \tilde x,z)  &=& \cos [u \tilde x] \cos [v \tilde x] (R_1(z)^2 + 
R_2(z)^2 uv \tilde x^2) + \sin [u \tilde x] \sin [v \tilde x] (R_1(z)^2 uv 
\tilde x^2+ R_2(z)^2 ) \nonumber \\
 &+& 2u \tilde x\{ R_1(z)^2  \cos [v \tilde x] \sin [u \tilde x]  -  R_2(z)^2 
\sin [v \tilde x] \cos [u \tilde x] \} \nonumber \\
 &+&  R_1(z) R_2(z)\{- 2u \tilde x \cos[ (v + u) \tilde x] + ( 1-  v u  \tilde 
x^2) \sin [ (v + u)\tilde x] \}. 
\earr

\section{Estimation of the tensor-to-scalar ratio}\label{estimation}

 The goal of this section is to obtain an estimation for the amplitude of the power 
spectrum. 

%
 
We begin by rearranging expression \eqref{psgw1}, this is

\beq\label{psgw2}
P_T (k,\eta) =  x^4 \frac{H^4}{8\pi^4M_P^4} \int_0^\infty dv 
\int_{|v-1|}^{|v+1|} du \frac{[4v^2 - (u^2-v^2 -1)^2]^2}{u^2v^2} 
I_{1}(v,u,x,z)I_{2}(v,u,x,z),
\eeq
where

\begin{subequations}
 \begin{gather}
I_{1}(v,u,x,z)=\int_{z}^{x} d\tilde{x_1} 
G(x,\tilde{x_1})F(v\tilde{x_1},u\tilde{x_1},z),\\
  I_{2}(v,u,x,z)=\int_{z}^{x} d\tilde{x_2} 
G(x,\tilde{x_2})F_2(v\tilde{x_2},u\tilde{x_2},z).
 \end{gather}

\end{subequations}
and $F_2(v\tilde{x},u\tilde{x},z)\equiv 
F(v\tilde{x},u\tilde{x},z)+F(u\tilde{x},v\tilde{x},z)$.

The tensor power spectrum obtained in Eq. \eqref{psgw2} is exact, no 
approximations have been made. On the other hand, we will be interested in the 
value of the power spectrum at a conformal time near the end of inflation, this 
is, at $x\equiv k\eta \to 0^-$. It is clear that the real measurement, say the 
amplitude of the $B$-modes  from CMB observations, is associated to the 
power spectrum evaluated at the time of decoupling. However, as we will show 
next, the amplitude of the power spectrum near the end of inflation is too low 
that it would be very hard to conceive that some physical process, occurring 
during the transition from the inflationary regime to the radiation dominated 
epoch, would amplify the power spectrum for several orders of magnitude in 
order to make it detectable by recent experiments.

We will consider two cases: i) the time of collapse occurs when the proper 
wavelength of the mode is larger than the Hubble radius, i.e. when $k < a(\tc) H(\tc)$ 
or equivalently when $-k \tc = |z| < 1$; and ii) the time of collapse occurs when the 
proper wavelength of the mode is smaller than the Hubble radius, i.e. when $k > a(\tc) 
H(\tc)$ or equivalently when $z = k \tc \to -\infty$. We will show that both cases lead 
to similar conclusions.

We start with the first case, that is, the time of collapse is such that $|z| < 1$ and 
 focus on the integral $I_1 (v,u,x,z)$,

\begin{equation}
I_{1}(v,u,x,z) = j_1(x)\int_{z}^{x} d\tilde{x_1} 
y_1(\tilde{x_1})F(v\tilde{x_1},u\tilde{x_1},z)-y_1(x)\int_{z}^{x} 
d\tilde{x_1} j_1(\tilde{x_1})F(v\tilde{x_1},u\tilde{x_1},z).
\label{I1a}
\end{equation}
Therefore, since we are interested in $x \to 0^-$ and since we are assuming that $|z| < 
1$, 
we can approximate

\barr\label{I1}
I_{1}(v,u,x,z) &\simeq& j_1(x)\int_{z}^{x} d\tilde{x_1} \bigg(   
\frac{-R_1(z)^2}{\tilde{x_1}^2} + \frac{R_1(z)R_2(z)}{\tilde{x_1}} (u-v) \bigg) - 
y_1(x)\int_{z}^{x}d\tilde{x_1} \bigg( \frac{R_1 (z)^2\tilde{x_1}}{3} + 
\frac{R_1(z)R_2(z)}{3} (v-u) \tilde{x_1}^2 \bigg) \nonumber \\
&\simeq& j_1(x) \bigg[R_1(z)^2 \bigg(\frac{1}{x} - \frac{1}{z} \bigg) + R_1(z) R_2(z) 
(v-u) \ln \frac{z}{x}   \bigg] \nn
&-& y_1(x) \bigg[\frac{R_1(z)}{6} (x^2-z^2) + 
\frac{R_1(z)R_2(z)}{9}(v-u)(x^3-z^3)  \bigg]\nonumber \\
&\simeq& \frac{ R_1(z)^2}{3} \bigg( \frac{3}{2} - \frac{x}{z} - \frac{z^2}{2x^2} \bigg) + 
\frac{R_1(z)R_2(z)}{3} (v-u) z \bigg(\frac{x}{z} \ln  \frac{z}{x} + 
\frac{x}{3z} - \frac{z^2}{3x^2} \bigg),
\earr
 where in the last line we have used the first non-vanishing term of the series 
for $j_1$ and $y_1$ (spherical Bessel's functions of 1st and 2nd kind of order 
1) when $x\to 0^-$. Turning our attention now to $I_2$, we have
 
 \begin{equation}
I_{2}(v,u,x,z)= j_1(x)\int_{z}^{x} d\tilde{x_2} 
y_1(\tilde{x_2})F_2(v\tilde{x_2},u\tilde{x_2})-y_1(x)\int_{z}^{x} 
d\tilde{x_2} j_1(\tilde{x_2})F_2(v\tilde{x_2},u\tilde{x_2}).
\label{I2a}
\end{equation}
 Once again,  by taking into account $x \to 0^-$, we can approximate

\barr\label{I2}
I_{2}(v,u,x,z) &\simeq& j_1(x)\int_{z}^{x}d\tilde{x_2}  \bigg(
\frac{-2R_1(z)^2}{\tilde{x_2}^2} -\frac{2}{\tilde{x_2}} (v-u) R_1(z) R_2(z)  \bigg) \nn
&-& 
y_1(x)\int_{z}^{x} d\tilde{x_2} \bigg(
\frac{2R_1(z)^2\tilde{x_2}}{3} +  \frac{2}{3} (v-u) R_1 (z) R_2(z) 
\tilde{x_2}^2\bigg)\nonumber \\
&\simeq& j_1(x) 2 \bigg[R_1(z)^2 \bigg(\frac{1}{x} - \frac{1}{z} \bigg) + R_1(z) R_2(z) 
(v-u) \ln \frac{z}{x} \bigg]  \nonumber \\
&-& y_1(x)\bigg[\frac{R_1(z)^2}{3} (x^2-z^2) + \frac{2}{9} R_1(z)R_2(z) (v-u) (x^3-z^3)   
 \bigg] \nn
 &\simeq& \frac{R_1(z)^3}{3} \bigg(3 - \frac{2x}{z}-\frac{z^2}{x^2} \bigg)  +\frac{2}{3} 
 R_1(z) R_2(z) (v-u) z \bigg(\frac{x}{z} \ln \frac{z}{x} + \frac{x}{3z} -\frac{z^2}{3x^2} 
\bigg) .
\earr

 
\begin{figure}
\begin{center}
\includegraphics[width=7.5cm,height=6.0cm]{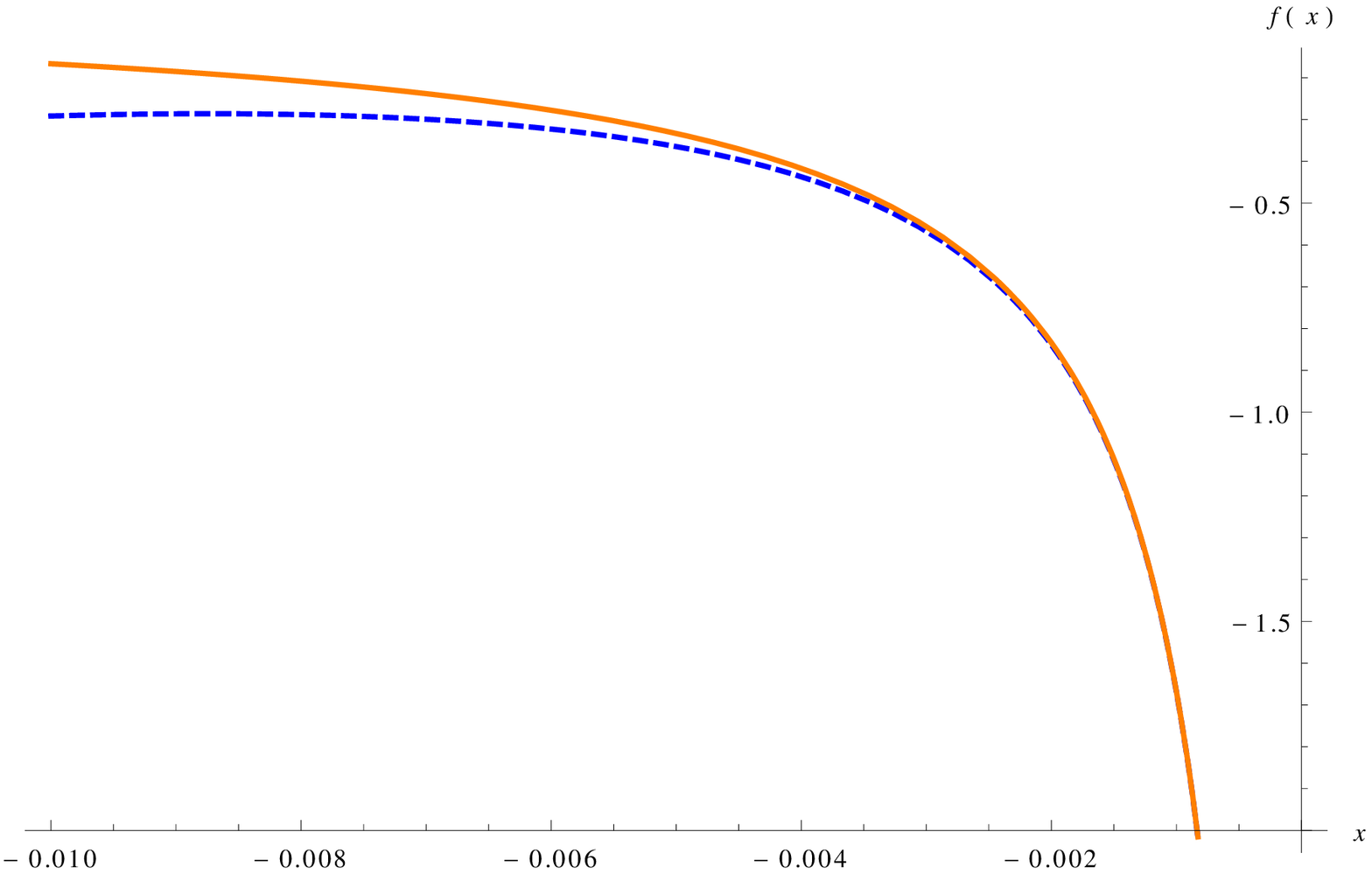} \label{f:grafico2}
\includegraphics[width=7.5cm,height=6.0cm]{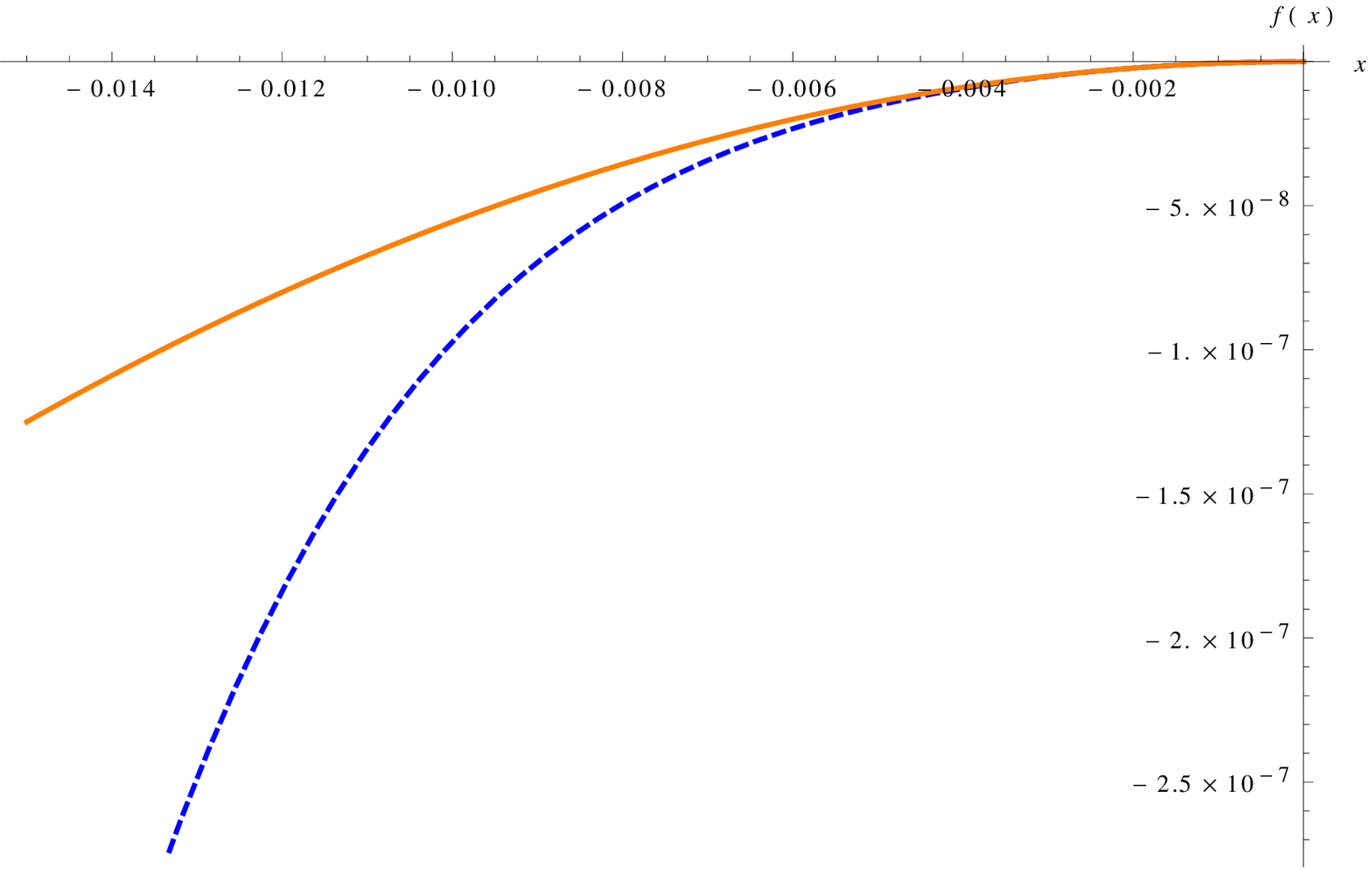}\label{f:grafico2a}
\end{center}
\caption{The left figure shows a plot of the first term of the integrand of Eq. 
\eqref{I1a} for fixed values  $u=1,v=0.5,z=-0.01$, i.e. $f(x)  \equiv 
y_1(x)F[0.5x,1x,-0.01]$ with $x \in [z,0]$. The right figure shows a plot of the second 
term of the integrand of Eq. \eqref{I1a} for fixed values  $u=1,v=0.5,z=-0.01$, i.e. $f(x) 
 \equiv j_1(x)F[0.5x,1x,-0.01]$ with $x \in [z,0]$
 The blue-dashed curves represent the total functions, while the orange ones represent the first two non-null terms of the  approximations. }
\label{f:graficos2}
\end{figure}
\begin{figure}
\begin{center}
\includegraphics[width=7.5cm,height=6.0cm]{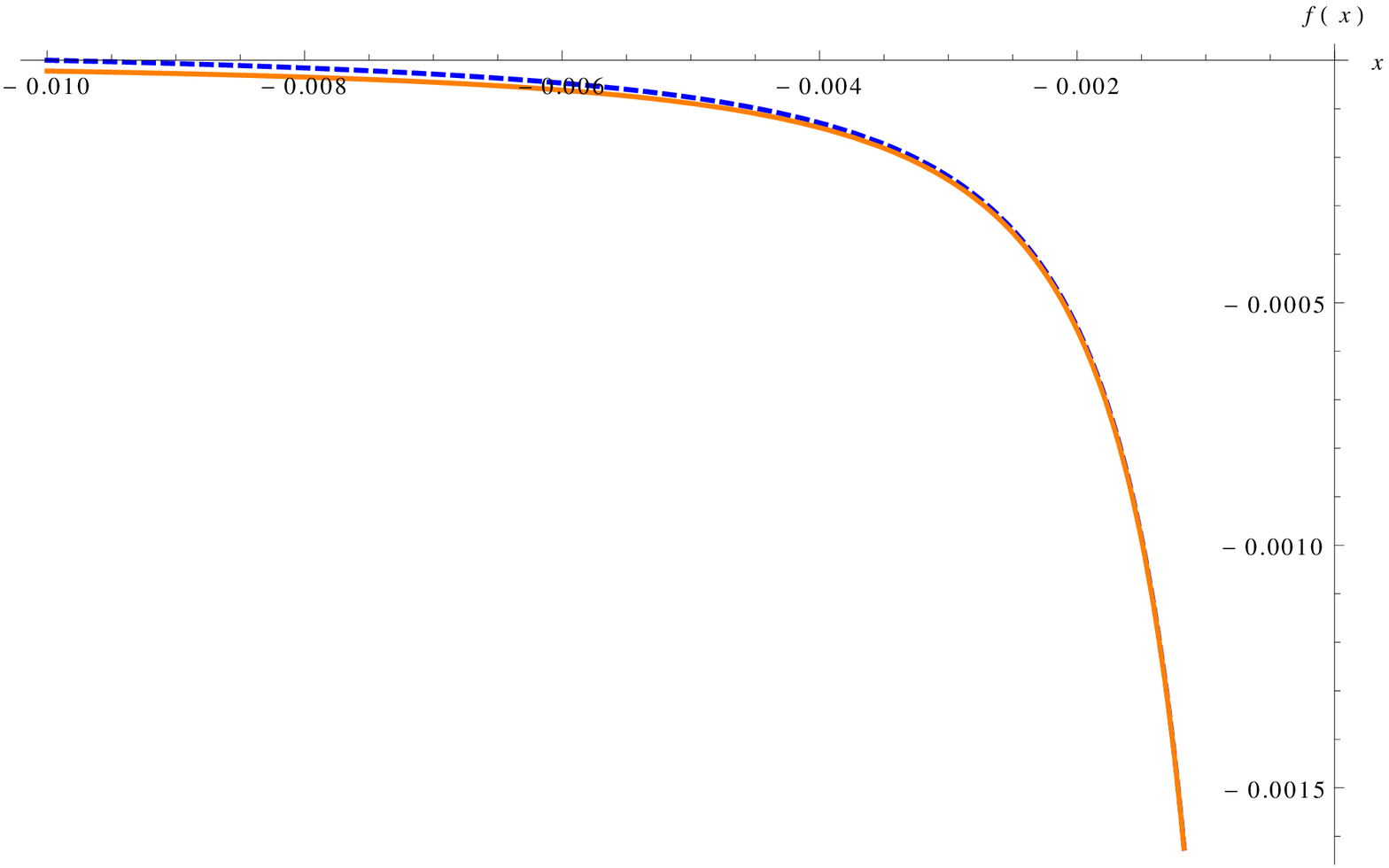}\label{f:grafico3}
\includegraphics[width=7.5cm,height=6.0cm]{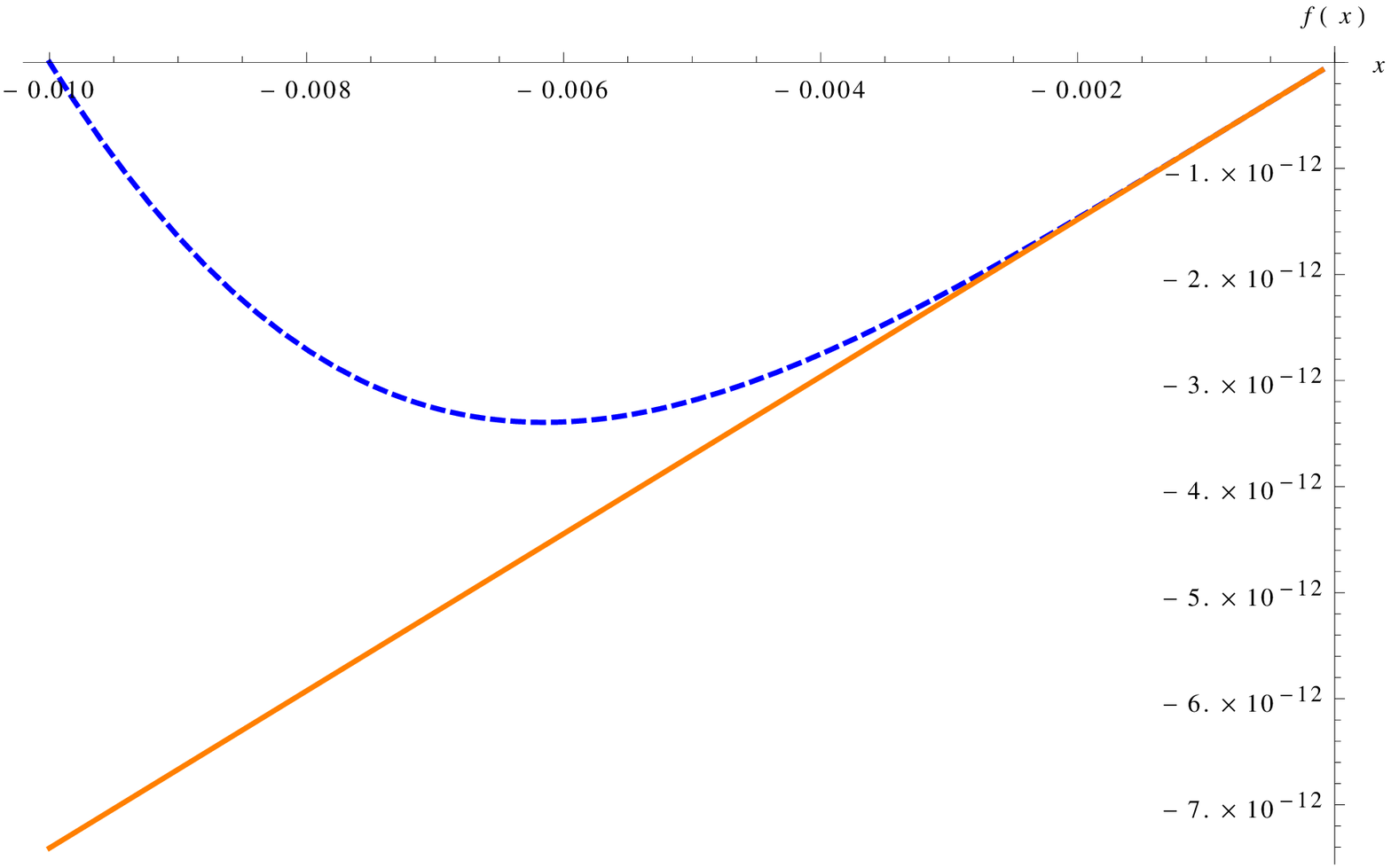}\label{f:grafico4}
\end{center} 
\caption{The left figure shows a plot of the first term of the integrand of Eq. 
 \eqref{I2a} for fixed values  $u=1,v=0.5,z=-0.01$, i.e. $f(x)  \equiv 
y_1(x)F_2[0.5x,1x,-0.01]$ with $x \in [z,0]$. The right figure shows a plot of 
the second term of the integrand of Eq. \eqref{I1a} for fixed values  
$u=1,v=0.5,z=-0.01$, i.e. $f(x)  \equiv j_1(x)F_2[0.5x,1x,-0.01]$ with $x \in [z,0]$
 The blue-dashed curves represent the total functions, while the orange ones represent the first two non-null terms of the  approximations.}
\label{f:graficos3y4}
\end{figure}

Figures (\ref{f:graficos2}) and (\ref{f:graficos3y4}) show that 
the approximations used in both integrals $I_{1}(v,u,x,z)$ and $I_{2}(v,u,x,z)$ 
are viable when $x\to 0^-$ and $u$, $v$ and $z$ are fixed (we have found that 
for different values of $v$, $u$ and $z$, the graphics are not significantly 
modified). Moreover $I_{2}(v,u,x,z) \simeq 2 I_{1}(v,u,x,z) $; consequently, 
substituting Eqs. \eqref{I1} and \eqref{I2} into Eq. 
\eqref{psgw1} yields
 
 \barr
 P_T (k,\eta) &\simeq&  x^4 \frac{H^4}{8\pi^4M_P^4}  \int_0^{\infty} dv 
\int_{|v-1|}^{|v+1|} du \frac{[4v^2 - (u^2-v^2-1)^2]^2}{u^2v^2} \nn
&\times& 2 \bigg[\frac{ R_1(z)^2}{3} \bigg( \frac{3}{2} - \frac{x}{z} - \frac{z^2}{2x^2} 
\bigg) + 
\frac{R_1(z)R_2(z)}{3} (v-u) z \bigg(\frac{x}{z} \ln  \frac{z}{x} + 
\frac{x}{3z} - \frac{z^2}{3x^2} \bigg) \bigg]^2, \nn
\earr

We can rewrite the above expression in the following way

\barr\label{PT}
P_T (k,\eta) &\simeq&  x^4 \frac{H^4}{8\pi^4M_P^4} \frac{2}{9} \bigg[ R_1(z)^4 \bigg( 
\frac{3}{2} - \frac{x}{z} - \frac{z^2}{2x^2} \bigg)^2 M_1 + 2 R_1(z)^3 R_2(z) z \bigg(  
\frac{3}{2} - \frac{x}{z} - \frac{z^2}{2x^2} \bigg) \bigg( \frac{x}{z} \ln \frac{z}{x} + 
\frac{x}{3z} - \frac{z^2}{3x^2} \bigg) M_2  \nn
&+& R_1(z)^2 R_2(z)^2 z^2 \bigg( \frac{x}{z} \ln \frac{z}{x} + 
\frac{x}{3z} - \frac{z^2}{3x^2} \bigg)^2 M_3\bigg] 
\earr

with $M_1$, $M_2$ and $M_3$ defined explicitly in Appendix \ref{appB}.

Furthermore, with the expressions for the tensor and scalar power spectra at hand,  Eqs. 
\eqref{PT} and  \eqref{PS} respectively, we can estimate the tensor-to-scalar ratio 
defined as $r \equiv P_T/P_S$, thus

\barr\label{r0}
r &\simeq& \frac{16 x^4 H^2 \epsilon }{9\pi^2 M_P^2}  \bigg[ R_1(z)^2 \bigg( 
\frac{3}{2} - \frac{x}{z} - \frac{z^2}{2x^2} \bigg)^2 M_1 + 2 R_1(z) R_2(z) z \bigg(  
\frac{3}{2} - \frac{x}{z} - \frac{z^2}{2x^2} \bigg) \bigg( \frac{x}{z} \ln \frac{z}{x} + 
\frac{x}{3z} - \frac{z^2}{3x^2} \bigg) M_2  \nn
&+&  R_2(z)^2 z^2 \bigg( \frac{x}{z} \ln \frac{z}{x} + 
\frac{x}{3z} - \frac{z^2}{3x^2} \bigg)^2 M_3\bigg] 
\earr
Note that in expression above, we have made use of Eq. \eqref{PS}, with $n_s \simeq 1$, 
i.e. $P_S(k,\eta) \simeq H^2/(64 \pi^2 M_P^2 \epsilon) R_1(z)^2$; also note that the 
function involving the time of collapse in the scalar power spectrum is exactly the same 
as $R_1(z)^2$.

Now we can use the fact that the observational data constrain the value of the scalar 
power spectrum to be $10^{-9}$. That is, in our approach, this constraint implies that 
$P_S(k,\eta) \simeq H^2/(64 \pi^2 M_P^2 \epsilon) R_1(z)^2 \simeq 10^{-9}$; thus,

\beq\label{H2}
\frac{H^2}{\pi^2 M_P^2} \simeq \frac{10^{-9} 8^2 \epsilon}{R_1(z)^2}.
\eeq
Equation \eqref{H2} means that, in order to our approach to be consistent with the 
observations associated to the scalar power spectrum, the time of collapse must satisfy 
$H^2/(64 \pi^2 M_P^2 \epsilon) R_1(z)^2 \simeq 10^{-9}$.

Substituting Eq. \eqref{H2} in Eq. \eqref{r0} yields

\barr\label{r1}
r &\simeq& \frac{1024}{9}10^{-9} \epsilon^2 x^4 \bigg[ \bigg( 
\frac{3}{2} - \frac{x}{z} - \frac{z^2}{2x^2} \bigg)^2 M_1 + 2 \frac{ R_2(z) z}{R_1(z)} 
\bigg(  
\frac{3}{2} - \frac{x}{z} - \frac{z^2}{2x^2} \bigg) \bigg( \frac{x}{z} \ln \frac{z}{x} + 
\frac{x}{3z} - \frac{z^2}{3x^2} \bigg) M_2  \nn
&+&  \frac{R_2(z)^2 z^2}{R_1(z)^2} \bigg( \frac{x}{z} \ln \frac{z}{x} + 
\frac{x}{3z} - \frac{z^2}{3x^2} \bigg)^2 M_3\bigg] 
\earr
Finally, since we have assumed that the time of collapse occurs when the 
proper wavelength of the mode is bigger than the Hubble radius, i.e. $ |z| = -k\tc <1$, 
then we can expand in series the functions $ R_2(z) z/R_1(z)$ and $R_2(z)^2 
z^2/R_1(z)^2$ around $z = 0$; additionally we will use  that $|z| = -k\tc > -k\eta = 
|x|$ with $x = k\eta \to 0^{-}$, and retain the leading terms in $x/z$. Thus, Eq. 
\eqref{r1} is approximated by

\beq\label{rfuera}
r\simeq \frac{1024}{9} 10^{-9} \epsilon^2 \bigg( \frac{z^4}{4} M_1 - z^2 M_2 + M_3  \bigg)
\eeq

Equation \eqref{rfuera} is the first main result of this section. We can see that the 
prediction for $r$ given by the collapse hypothesis within the semiclassical gravity 
framework, is suppressed by a factor of $10^{-9} \epsilon^2$ and by powers of $|z|<1$. In 
other words, the generically predicted tensor-to-scalar ratio is extremely small making 
it practically undetectable by recent or future experiments. However, it is still within 
the observational bounds provided by the latest \emph{Planck} results 
\cite{PlanckBicep15,Planckinflation15}, which set $r < 0.12$. Moreover, Eq. \eqref{rfuera} shows a degeneration between the slow roll parameter $\epsilon$ and the time of collapse.
Therefore, if future experiments confirm a possible detection of primordial 
gravitational waves, in clear contrast  to the standard approach, the measured value of $r$ would 
not fix the value of the slow-roll parameter $\epsilon$, it would only fix a combination of $\epsilon$ and the value of 
the time of collapse $z$.  Consequently, the confirmed detection 
of primordial $B$-modes would make this whole scenario, i.e. the self-induced collapse 
hypothesis within the semiclassical gravity, extremely unlikely.

We will focus now on the second case, that is,  the time of collapse occurs when 
the proper wavelength of the mode is smaller than the Hubble radius $z=k\tc \to -\infty$ 
i.e. at the very beginning of the inflationary regime. 

We begin this case with  Eq. \eqref{psgw2} which corresponds to the tensor power spectrum 
and proceed to calculate the tensor-to-scalar ratio $r \equiv P_T/P_S$ where $P_S(k,\eta) 
\simeq H^2/(64 \pi^2 M_P^2 \epsilon )R_1(z)^2$, Eq. \eqref{PS}; additionally, by 
considering that the time of collapse must satisfy $H^2/(64 \pi^2 M_P^2 \epsilon) 
R_1(z)^2 
\simeq 10^{-9}$, the prediction for $r$ is

\beq\label{rdentro1}
r \simeq 8^3 10^{-9} \epsilon^2 \frac{x^4}{R_1(z)^4} \int_0^\infty dv 
\int_{|v-1|}^{|v+1|} du \frac{[4v^2 - (u^2-v^2 -1)^2]^2}{u^2v^2} 
I_{1}(v,u,x,z)I_{2}(v,u,x,z)
\eeq
Next, we focus on the integrals $I_{1}(v,u,x,z),I_{2}(v,u,x,z)$. These integrals can be 
rewritten as

\begin{subequations}\label{int12}
\beq
I_{1}(v,u,x,z) = \int_{-\infty}^{x} d\tilde{x_1} G(x,\tilde{x_1})[R_1(z)^2 f_1(v 
\tilde{x_1}, u \tilde{x_1}) + R_1(z)R_2(z) f_2(v \tilde{x_1}, u \tilde{x_1}) + R_2(z)^2 
f_3(v\tilde{x_1}, u \tilde{x_1})],
\eeq
\beq
I_{2}(v,u,x,z) = \int_{-\infty}^{x} d\tilde{x_2} G(x,\tilde{x_2})[R_1(z)^2 \tilde f_1(v 
\tilde{x_2}, u \tilde{x_2}) + R_1(z)R_2(z) \tilde f_2(v \tilde{x_2}, u \tilde{x_2}) + 
R_2(z)^2 \tilde f_3(v\tilde{x_2}, u \tilde{x_2})],
\eeq
\end{subequations}
note that the lower limit of integration takes into account that  we are considering the 
case $z \to -\infty$. The explicit forms of the functions $f_1,f_2,f_3$ and 
$\tilde f_1, \tilde f_2,\tilde f_3$ can be found in Appendix \ref{appB}. The advantage of 
writing the integrals $I_{1}(v,u,x,z),I_{2}(v,u,x,z)$ in the form of Eqs. \eqref{int12} 
is that the dependence on the time of collapse is totally contained in the functions 
$R_1(z)$ and $R_2(z)$. 

Substituting Eqs. \eqref{int12} in Eq. \eqref{rdentro1} we have

\beq\label{rdentro2}
r\simeq 10^{-9} \epsilon^2 \bigg[  C_0(x) + C_1(x) \frac{R_2(z)}{R_1(z)} + C_2(x) \bigg( 
\frac{R_2(z)}{R_1(z)} \bigg)^2 + C_3(x) \bigg( 
\frac{R_2(z)}{R_1(z)} \bigg)^3 + C_4 (x) \bigg( 
\frac{R_2(z)}{R_1(z)} \bigg)^4      \bigg],
\eeq
note that $r$ written in this form exhibits explicitly the dependence on the time of 
collapse $z =k\tc$. The explicit form of the functions 
$C_0(x),C_1(x),C_2(x),C_3(x),C_4(x)$ can be found in Appendix \ref{appB}; also, it should 
be noted that such functions should be evaluated at $x \to 0^{-}$ i.e. at a time near the 
end of inflation. 

Next, using that $R_2(z)/R_1(z) = (\sin z + \cos z/z )/(\cos z - \sin z/z)$ and since we 
are interested in the case $z \to -\infty$, then we can approximate 

\beq
\frac{R_2(z)}{R_1(z)} \simeq \tan(z) \equiv \zeta,
\eeq
hence $-\infty < \zeta < \infty$. consequently Eq. \eqref{rdentro2} is given by

\beq\label{rdentro}
r \simeq 10^{-9} \epsilon^2 \bigg[  C_0(x) + C_1(x) \zeta + C_2(x) \zeta^2 + C_3(x) 
\zeta^3 + C_4 (x) \zeta^4  \bigg] \qquad \textrm{with} \qquad x\to 0^{-}.
\eeq

Equation \eqref{rdentro} is the second main result of this section. It shows the 
prediction for $r$ in the case when the time of collapse occurs near the beginning of 
the inflationary era. As in the previous case, 
the generic prediction for $r$, within our approach, is suppressed by a factor of 
$10^{-9} \epsilon^2$. Note also that, even if $\zeta$ is in the range  
$(-\infty,\infty)$, its value starts to grow only near  $z = -n \pi /2$ [with $n$ a 
(large) odd number], that is, for a generic value of $z$, the function $\zeta = \tan(z)$ 
does not grows arbitrarily large. Therefore, the prediction for $r$ in this case is also 
 consistent with recent observational data given that it is suppressed by the square of 
the slow-roll parameter. 

As in the case when $|z| < 1$, a hypothetical measurement of $r$ would translate into a bound  on a combination of the slow-roll parameter $\epsilon$ and the time of collapse. Consequently, the same discussion about the plausibility of the collapse model in the case when the time 
of collapse satisfies $-k\tc <1$ also applies in the present case, i.e. when the time of 
collapse is such that $k\tc \to -\infty$.

To end this section, let us summarize the general conditions and motivations under 
which results \eqref{rfuera} and \eqref{rdentro} were achieved and their extension to 
specific 
inflationary models. 

Our collapse proposal is meant to serve as a possible 
explanation to the generation of primordial perturbations, in particular the 
primordial gravitational waves. In a broad sense, we are considering the 
standard inflationary paradigm with the addition of the self-induced collapse hypothesis 
and the semiclassical gravity framework. In the present article, we have focused on the 
simplest inflationary model (and also the one favored by recent Planck data  
\cite{Planckinflation15}), namely a single adiabatic Gaussian scalar field (the inflaton) 
minimally coupled to gravity with canonical kinetic  term in the slow-roll 
approximation.

As regards with the characteristics of the self-induced collapse, the only assumption we 
have made is  that the time of collapse $\tc$ is proportional to $k^{-1}$; this 
relation was phenomenologically inferred from previous works by 
comparing the theoretical predictions (specifically the scalar angular power 
spectrum) with the observational data. Therefore, the relation $ k \tc$ 
implies a specific dependence of the time of collapse: $\tc = 
z/k$, being $z$ a free parameter of the collapse model.

In fact, any inflationary model satisfying the previous 
conditions (with the inclusion of the self-induced collapse and the 
semiclassical Einstein equations) would yield a similar suppression for  $r$ 
that is compatible with results \eqref{rfuera} and \eqref{rdentro}. For example, 
Starobinsky's inflationary model \cite{starobinsky,mukhanov81,starobinsky2} 
(also known as $R+R^2$ inflation) can be characterized by the action of a 
single-scalar field with canonical kinetic term, and a potential with a region 
in which the slow-roll approximation is valid; Starobinsky's model 
predicts a value \cite{Kehagias2013} for the 
slow-roll parameter  $|\epsilon| \simeq 3/(4N^2)$, with $N$ the number of 
$e$-foldings characteristic of inflation; thus, for $N \simeq 60$, one has 
$|\epsilon| \simeq 10^{-4}$. This is, for the Starobinsky's model, the value of 
$r$, within our approach, would be suppressed by a factor $10^{-9} \epsilon^2 =  10^{-17}$
Thus, the conclusions drawn from the results \eqref{rfuera} and \eqref{rdentro} apply to 
generic single-field slow-roll models within the semiclassical gravity framework and the 
self-induced collapse.

\section{Summary and Conclusions}
\label{conclusions}

In this paper we have computed the tensor primordial power spectrum and the 
tensor-to-scalar ratio $r$ by making use of semiclassical Einstein equations 
and also including a self-induced collapse of the inflaton's wave function. 
Within the semiclassical gravity approximation there is no source for tensor 
modes at the first-order perturbation theory; to calculate the tensor power spectrum the 
second-order has to be considered.

We have considered two cases: 1) the time of collapse occurs when the 
proper wavelength of the mode is \emph{greater} than the Hubble radius, that is $-k\tc < 
1$ and 2) the time of collapse occurs when the proper wavelength of the mode is 
\emph{smaller} than the Hubble radius, that is $k\tc \to -\infty$. In both cases a 
generic value for $z$ results in a prediction of $r$ that is suppressed by a factor of 
$10^{-9} \epsilon^2$, that is, a very small value; nevertheless, it is still consistent 
 with the limit obtained by the joint analysis performed by the BICEP/KECK and PLANCK 
collaborations. 

On the contrary, if a detection of primordial $B$ modes is confirmed, then the 
self-induced collapse hypothesis plus the semiclassical gravity approach would face 
severe issues given the degeneration between the slow roll parameter $\epsilon$ and the 
time of collapse in the prediction of $r$ [see Eqs.  \eqref{rfuera} and \eqref{rdentro}].

However, we have also 
discussed in this paper, that 
the discrepancy with standard inflationary model prediction's arises in considering the semiclassical approximation and not 
the self-induced collapse of the inflaton's wave function. Within the 
semiclassical approximation, tensor modes are different from zero only at 
second-order perturbation theory, while in the standard procedure (where a joint 
quantization of the metric and field is performed) the tensor modes can be 
generated at first-order in the perturbation theory. Furthermore, a recent 
calculation of the tensor-to-scalar ratio using the Mukhanov-Sasaki variable and 
including the collapse of the wave function \cite{MBL14,LB15} gives a 
prediction of the value of $r$ closest to the upper bound given by the joint 
BICEP/KECK and Planck collaboration's analysis.

 \acknowledgments
 
The authors thank D. Sudarsky and G. R. Bengochea for meaningful discussions. 
Support for this work was provided by PIP 11220120100504/14 CONICET and UNLP 
G119. G.L.'s research funded by Consejo Nacional de Investigaciones 
Cient\'{i}ficas y T\'{e}cnicas (CONICET), Argentina, and Consejo Nacional de 
Ciencia y Tecnolog\'{i}a (CONACYT), Mexico. We also thank the referees for 
useful suggestions. 
 
\appendix

\section{Expressions used in Sec. \ref{estimation}}\label{appB}

In this appendix we provide the explicit form of the expression used in Sec. 
\ref{estimation}.

The quantities $M_1$, $M_2$ and $M_3$ are defined as

\begin{subequations}
 \beq
 M_1 \equiv \int_0^{\infty} dv \int_{|v-1|}^{|v+1|} du \frac{[4v^2 - 
(u^2-v^2-1)^2]^2}{u^2v^2},
 \eeq
 \beq
  M_2 \equiv \int_0^{\infty} dv \int_{|v-1|}^{|v+1|} du \frac{[4v^2 - 
(u^2-v^2-1)^2]^2}{u^2v^2} (v-u),
 \eeq
 \beq
   M_3 \equiv \int_0^{\infty} dv \int_{|v-1|}^{|v+1|} du \frac{[4v^2 - 
(u^2-v^2-1)^2]^2}{u^2v^2}(v-u)^2.
\eeq
\end{subequations}

The functions $f_1(v \tilde{x_1},u\tilde{x_1})$, $f_2(v \tilde{x_1},u\tilde{x_1})$ 
and $f_3(v \tilde{x_1},u\tilde{x_1})$ used in Eqs. \eqref{int12} are defined as

\begin{subequations}
 \beq
 f_1(v \tilde{x_1},u\tilde{x_1}) \equiv \cos(u\tilde{x_1}) \cos(v\tilde{x_1}) + 
uv\tilde{x_1}^2 \sin(u \tilde{x_1})\sin (v\tilde{x_1}) + 2u\tilde{x_1} \cos(v\tilde{x_1}) 
\sin (u\tilde{x_1}),
 \eeq
 \beq
  f_2(v \tilde{x_1},u\tilde{x_1}) \equiv -2 u\tilde{x_1} \cos[(u+v)\tilde{x_1}]  + 
(1-uv\tilde{x_1}^2) \sin[(u+v)\tilde{x_1}],
 \eeq
 \beq
  f_3(v \tilde{x_1},u\tilde{x_1}) \equiv uv \tilde{x_1}^2 \cos (u\tilde{x_1}) 
\cos(v\tilde{x_1})+ \sin(u\tilde{x_1}) \sin (v\tilde{x_1}) 
-2u\tilde{x_1}\sin(v\tilde{x_1}) \cos(u\tilde{x_1}).
 \eeq
\end{subequations}

Meanwhile, the functions $\tilde f_1(v \tilde{x_1},u\tilde{x_1})$, $\tilde f_2(v 
\tilde{x_1},u\tilde{x_1})$ and $\tilde f_3(v \tilde{x_1},u\tilde{x_1})$ also  used 
in Eqs. \eqref{int12} are defined as
\begin{subequations}
 \beq
\tilde f_1(v \tilde{x_2},u\tilde{x_2}) \equiv 2[\cos(u\tilde{x_2}) \cos(v\tilde{x_2}) + 
uv\tilde{x_2}^2 \sin(u \tilde{x_2})\sin (v\tilde{x_2}) + 2u\tilde{x_2} \cos(v\tilde{x_2}) 
\sin (u\tilde{x_2}) + 2v\tilde{x_2} \cos (u\tilde{x_2}) \sin(v\tilde{x_2})],
 \eeq
 \beq
\tilde  f_2(v \tilde{x_2},u\tilde{x_2}) \equiv (-2 u\tilde{x_2}-2v\tilde{x_2}) 
\cos[(u+v)\tilde{x_2}]  + 2 (1-uv\tilde{x_2}^2) \sin[(u+v)\tilde{x_2}],
 \eeq
 \beq
\tilde  f_3(v \tilde{x_2},u\tilde{x_2}) \equiv 2[ uv \tilde{x_2}^2 \cos (u\tilde{x_2}) 
\cos(v\tilde{x_2})+ \sin(u\tilde{x_2}) \sin (v\tilde{x_2}) 
-u\tilde{x_2}\sin(v\tilde{x_2}) \cos(u\tilde{x_2})-v\tilde{x_2} \cos (v\tilde{x_2}).
\sin(u\tilde{x_2})]
 \eeq
\end{subequations}

The functions $C_0(x)$, $C_1(x)$, $C_2(x)$, $C_3(x)$ and $C_4(x)$ used in Eqs. 
\eqref{rdentro2} and \eqref{rdentro} are defined as

\begin{subequations}
 \barr
 C_0(x) &\equiv& 8^3 x^4 \int_0^\infty dv \int_{|v-1|}^{|v+1|} du \int_{-\infty}^x d 
 \tilde{x_1} \int_{-\infty}^x d \tilde{x_2} \: G(x,\tilde x_1) G(x,\tilde x_2) 
\frac{[4v^2 - (u^2-v^2-1)^2]^2}{u^2v^2} \nn
&\times& f_1(v \tilde{x_1},u\tilde{x_1})  \tilde f_1(v \tilde{x_2},u\tilde{x_2}),
 \earr
  \barr
 C_1(x) &\equiv& 8^3 x^4 \int_0^\infty dv \int_{|v-1|}^{|v+1|} du \int_{-\infty}^x d 
 \tilde{x_1} \int_{-\infty}^x d \tilde{x_2} \:  G(x,\tilde x_1) G(x,\tilde x_2) 
\frac{[4v^2 -(u^2-v^2-1)^2]^2}{u^2v^2} \nn
&\times& [ f_2(v \tilde{x_1},u\tilde{x_1})  \tilde f_1(v \tilde{x_2},u\tilde{x_2}) + 
f_1(v \tilde{x_1},u\tilde{x_1})  \tilde f_2(v \tilde{x_2},u\tilde{x_2}) ],
 \earr
  \barr
 C_2(x) &\equiv& 8^3 x^4 \int_0^\infty dv \int_{|v-1|}^{|v+1|} du \int_{-\infty}^x d 
 \tilde{x_1} \int_{-\infty}^x d \tilde{x_2} \:  G(x,\tilde x_1) G(x,\tilde x_2) 
\frac{[4v^2 - (u^2-v^2-1)^2]^2}{u^2v^2} \nn
&\times& [f_3(v \tilde{x_1},u\tilde{x_1})  \tilde f_1(v \tilde{x_2},u\tilde{x_2}) + f_1(v 
\tilde{x_1},u\tilde{x_1})  \tilde f_3(v \tilde{x_2},u\tilde{x_2}) + f_2(v 
\tilde{x_1},u\tilde{x_1})  \tilde f_2(v \tilde{x_2},u\tilde{x_2}) ],
 \earr
  \barr
 C_3(x) &\equiv& 8^3 x^4 \int_0^\infty dv \int_{|v-1|}^{|v+1|} du \int_{-\infty}^x d 
 \tilde{x_1} \int_{-\infty}^x d \tilde{x_2} \:  G(x,\tilde x_1) G(x,\tilde x_2) 
\frac{[4v^2 -(u^2-v^2-1)^2]^2}{u^2v^2} \nn
&\times&  [ f_3(v \tilde{x_1},u\tilde{x_1})  \tilde f_2(v \tilde{x_2},u\tilde{x_2}) + 
f_2(v \tilde{x_1},u\tilde{x_1})  \tilde f_3(v \tilde{x_2},u\tilde{x_2}) ],
 \earr
 \barr
 C_4(x) &\equiv& 8^3 x^4 \int_0^\infty dv \int_{|v-1|}^{|v+1|} du \int_{-\infty}^x d 
 \tilde{x_1} \int_{-\infty}^x d \tilde{x_2} \:  G(x,\tilde x_1) G(x,\tilde x_2) 
\frac{[4v^2 -(u^2-v^2-1)^2]^2}{u^2v^2} \nn
&\times& f_3(v \tilde{x_1},u\tilde{x_1})  \tilde f_3(v \tilde{x_2},u\tilde{x_2}) .
 \earr
\end{subequations}

\bibliography{bibliografia}

\begin{thebibliography}{49}
\expandafter\ifx\csname natexlab\endcsname\relax\def\natexlab#1{#1}\fi
\expandafter\ifx\csname bibnamefont\endcsname\relax
  \def\bibnamefont#1{#1}\fi
\expandafter\ifx\csname bibfnamefont\endcsname\relax
  \def\bibfnamefont#1{#1}\fi
\expandafter\ifx\csname citenamefont\endcsname\relax
  \def\citenamefont#1{#1}\fi
\expandafter\ifx\csname url\endcsname\relax
  \def\url#1{\texttt{#1}}\fi
\expandafter\ifx\csname urlprefix\endcsname\relax\def\urlprefix{URL }\fi
\providecommand{\bibinfo}[2]{#2}
\providecommand{\eprint}[2][]{\url{#2}}

\bibitem[{\citenamefont{{Martin}
  et~al.}(2014{\natexlab{a}})\citenamefont{{Martin}, {Ringeval}, {Trotta}, and
  {Vennin}}}]{Martin14}
\bibinfo{author}{\bibfnamefont{J.}~\bibnamefont{{Martin}}},
  \bibinfo{author}{\bibfnamefont{C.}~\bibnamefont{{Ringeval}}},
  \bibinfo{author}{\bibfnamefont{R.}~\bibnamefont{{Trotta}}}, \bibnamefont{and}
  \bibinfo{author}{\bibfnamefont{V.}~\bibnamefont{{Vennin}}},
  \bibinfo{journal}{Journal of Cosmology and Astroparticle Physics}
  \textbf{\bibinfo{volume}{3}}, \bibinfo{eid}{039}
  (\bibinfo{year}{2014}{\natexlab{a}}), \eprint{1312.3529}.

\bibitem[{\citenamefont{{Martin}
  et~al.}(2014{\natexlab{b}})\citenamefont{{Martin}, {Ringeval}, and
  {Vennin}}}]{infmodels}
\bibinfo{author}{\bibfnamefont{J.}~\bibnamefont{{Martin}}},
  \bibinfo{author}{\bibfnamefont{C.}~\bibnamefont{{Ringeval}}},
  \bibnamefont{and} \bibinfo{author}{\bibfnamefont{V.}~\bibnamefont{{Vennin}}},
  \bibinfo{journal}{Physics of the Dark Universe} \textbf{\bibinfo{volume}{5}},
  \bibinfo{pages}{75} (\bibinfo{year}{2014}{\natexlab{b}}), \eprint{1303.3787}.

\bibitem[{\citenamefont{{The Polarbear Collaboration: P.~A.~R.~Ade}
  et~al.}(2014)\citenamefont{{The Polarbear Collaboration: P.~A.~R.~Ade},
  {Akiba}, {Anthony}, {Arnold}, {Atlas}, {Barron}, {Boettger}, {Borrill},
  {Chapman}, {Chinone} et~al.}}]{polarbear14}
\bibinfo{author}{\bibnamefont{{The Polarbear Collaboration: P.~A.~R.~Ade}}},
  \bibinfo{author}{\bibfnamefont{Y.}~\bibnamefont{{Akiba}}},
  \bibinfo{author}{\bibfnamefont{A.~E.} \bibnamefont{{Anthony}}},
  \bibinfo{author}{\bibfnamefont{K.}~\bibnamefont{{Arnold}}},
  \bibinfo{author}{\bibfnamefont{M.}~\bibnamefont{{Atlas}}},
  \bibinfo{author}{\bibfnamefont{D.}~\bibnamefont{{Barron}}},
  \bibinfo{author}{\bibfnamefont{D.}~\bibnamefont{{Boettger}}},
  \bibinfo{author}{\bibfnamefont{J.}~\bibnamefont{{Borrill}}},
  \bibinfo{author}{\bibfnamefont{S.}~\bibnamefont{{Chapman}}},
  \bibinfo{author}{\bibfnamefont{Y.}~\bibnamefont{{Chinone}}},
  \bibnamefont{et~al.}, \bibinfo{journal}{\apj} \textbf{\bibinfo{volume}{794}},
  \bibinfo{eid}{171} (\bibinfo{year}{2014}), \eprint{1403.2369}.

\bibitem[{\citenamefont{{Ade} et~al.}(2014)\citenamefont{{Ade}, {Aikin},
  {Barkats}, {Benton}, {Bischoff}, {Bock}, {Brevik}, {Buder}, {Bullock},
  {Dowell} et~al.}}]{BICEP2}
\bibinfo{author}{\bibfnamefont{P.~A.~R.} \bibnamefont{{Ade}}},
  \bibinfo{author}{\bibfnamefont{R.~W.} \bibnamefont{{Aikin}}},
  \bibinfo{author}{\bibfnamefont{D.}~\bibnamefont{{Barkats}}},
  \bibinfo{author}{\bibfnamefont{S.~J.} \bibnamefont{{Benton}}},
  \bibinfo{author}{\bibfnamefont{C.~A.} \bibnamefont{{Bischoff}}},
  \bibinfo{author}{\bibfnamefont{J.~J.} \bibnamefont{{Bock}}},
  \bibinfo{author}{\bibfnamefont{J.~A.} \bibnamefont{{Brevik}}},
  \bibinfo{author}{\bibfnamefont{I.}~\bibnamefont{{Buder}}},
  \bibinfo{author}{\bibfnamefont{E.}~\bibnamefont{{Bullock}}},
  \bibinfo{author}{\bibfnamefont{C.~D.} \bibnamefont{{Dowell}}},
  \bibnamefont{et~al.}, \bibinfo{journal}{Physical Review Letters}
  \textbf{\bibinfo{volume}{112}}, \bibinfo{eid}{241101} (\bibinfo{year}{2014}),
  \eprint{1403.3985}.

\bibitem[{\citenamefont{{Mortonson} and {Seljak}}(2014)}]{Mortonson14}
\bibinfo{author}{\bibfnamefont{M.~J.} \bibnamefont{{Mortonson}}}
  \bibnamefont{and} \bibinfo{author}{\bibfnamefont{U.}~\bibnamefont{{Seljak}}},
  \bibinfo{journal}{Journal of Cosmology and Astroparticle Physics}
  \textbf{\bibinfo{volume}{10}}, \bibinfo{eid}{035} (\bibinfo{year}{2014}),
  \eprint{1405.5857}.

\bibitem[{\citenamefont{Flauger et~al.}(2014)\citenamefont{Flauger, Hill, and
  Spergel}}]{Flauger14}
\bibinfo{author}{\bibfnamefont{R.}~\bibnamefont{Flauger}},
  \bibinfo{author}{\bibfnamefont{J.~C.} \bibnamefont{Hill}}, \bibnamefont{and}
  \bibinfo{author}{\bibfnamefont{D.~N.} \bibnamefont{Spergel}},
  \bibinfo{journal}{JCAP} \textbf{\bibinfo{volume}{1408}}, \bibinfo{pages}{039}
  (\bibinfo{year}{2014}), \eprint{1405.7351}.

\bibitem[{\citenamefont{{Liu} et~al.}(2014)\citenamefont{{Liu}, {Mertsch}, and
  {Sarkar}}}]{Liu14}
\bibinfo{author}{\bibfnamefont{H.}~\bibnamefont{{Liu}}},
  \bibinfo{author}{\bibfnamefont{P.}~\bibnamefont{{Mertsch}}},
  \bibnamefont{and} \bibinfo{author}{\bibfnamefont{S.}~\bibnamefont{{Sarkar}}},
  \bibinfo{journal}{Astrophysical Journal Letters}
  \textbf{\bibinfo{volume}{789}}, \bibinfo{eid}{L29} (\bibinfo{year}{2014}),
  \eprint{1404.1899}.

\bibitem[{\citenamefont{{Planck Collaboration}
  et~al.}(2014)\citenamefont{{Planck Collaboration}, {Adam}, {Ade}, {Aghanim},
  {Arnaud}, {Aumont}, {Baccigalupi}, {Banday}, {Barreiro}, {Bartlett}
  et~al.}}]{planckdust}
\bibinfo{author}{\bibnamefont{{Planck Collaboration}}},
  \bibinfo{author}{\bibfnamefont{R.}~\bibnamefont{{Adam}}},
  \bibinfo{author}{\bibfnamefont{P.~A.~R.} \bibnamefont{{Ade}}},
  \bibinfo{author}{\bibfnamefont{N.}~\bibnamefont{{Aghanim}}},
  \bibinfo{author}{\bibfnamefont{M.}~\bibnamefont{{Arnaud}}},
  \bibinfo{author}{\bibfnamefont{J.}~\bibnamefont{{Aumont}}},
  \bibinfo{author}{\bibfnamefont{C.}~\bibnamefont{{Baccigalupi}}},
  \bibinfo{author}{\bibfnamefont{A.~J.} \bibnamefont{{Banday}}},
  \bibinfo{author}{\bibfnamefont{R.~B.} \bibnamefont{{Barreiro}}},
  \bibinfo{author}{\bibfnamefont{J.~G.} \bibnamefont{{Bartlett}}},
  \bibnamefont{et~al.}, \bibinfo{journal}{ArXiv e-prints}
  (\bibinfo{year}{2014}), \eprint{1409.5738}.

\bibitem[{\citenamefont{{BICEP2/Keck and Planck Collaborations}
  et~al.}(2015)\citenamefont{{BICEP2/Keck and Planck Collaborations}, {Ade},
  {Aghanim}, {Ahmed}, {Aikin}, {Alexander}, {Arnaud}, {Aumont}, {Baccigalupi},
  {Banday} et~al.}}]{PlanckBicep15}
\bibinfo{author}{\bibnamefont{{BICEP2/Keck and Planck Collaborations}}},
  \bibinfo{author}{\bibfnamefont{P.~A.~R.} \bibnamefont{{Ade}}},
  \bibinfo{author}{\bibfnamefont{N.}~\bibnamefont{{Aghanim}}},
  \bibinfo{author}{\bibfnamefont{Z.}~\bibnamefont{{Ahmed}}},
  \bibinfo{author}{\bibfnamefont{R.~W.} \bibnamefont{{Aikin}}},
  \bibinfo{author}{\bibfnamefont{K.~D.} \bibnamefont{{Alexander}}},
  \bibinfo{author}{\bibfnamefont{M.}~\bibnamefont{{Arnaud}}},
  \bibinfo{author}{\bibfnamefont{J.}~\bibnamefont{{Aumont}}},
  \bibinfo{author}{\bibfnamefont{C.}~\bibnamefont{{Baccigalupi}}},
  \bibinfo{author}{\bibfnamefont{A.~J.} \bibnamefont{{Banday}}},
  \bibnamefont{et~al.}, \bibinfo{journal}{Physical Review Letters}
  \textbf{\bibinfo{volume}{114}}, \bibinfo{eid}{101301} (\bibinfo{year}{2015}),
  \eprint{1502.00612}.

\bibitem[{\citenamefont{{Landau} et~al.}(2013)\citenamefont{{Landau},
  {Le{\'o}n}, and {Sudarsky}}}]{LLS13}
\bibinfo{author}{\bibfnamefont{S.~J.} \bibnamefont{{Landau}}},
  \bibinfo{author}{\bibfnamefont{G.}~\bibnamefont{{Le{\'o}n}}},
  \bibnamefont{and}
  \bibinfo{author}{\bibfnamefont{D.}~\bibnamefont{{Sudarsky}}},
  \bibinfo{journal}{Physical Review D} \textbf{\bibinfo{volume}{88}},
  \bibinfo{eid}{023526} (\bibinfo{year}{2013}), \eprint{1107.3054}.

\bibitem[{\citenamefont{Perez et~al.}(2006)\citenamefont{Perez, Sahlmann, and
  Sudarsky}}]{PSS06}
\bibinfo{author}{\bibfnamefont{A.}~\bibnamefont{Perez}},
  \bibinfo{author}{\bibfnamefont{H.}~\bibnamefont{Sahlmann}}, \bibnamefont{and}
  \bibinfo{author}{\bibfnamefont{D.}~\bibnamefont{Sudarsky}},
  \bibinfo{journal}{Class. Quant. Grav.} \textbf{\bibinfo{volume}{23}},
  \bibinfo{pages}{2317} (\bibinfo{year}{2006}), \eprint{gr-qc/0508100}.

\bibitem[{\citenamefont{{Sudarsky}}(2011)}]{Shortcomings}
\bibinfo{author}{\bibfnamefont{D.}~\bibnamefont{{Sudarsky}}},
  \bibinfo{journal}{International Journal of Modern Physics D}
  \textbf{\bibinfo{volume}{20}}, \bibinfo{pages}{509} (\bibinfo{year}{2011}),
  \eprint{0906.0315}.

\bibitem[{\citenamefont{Sudarsky}(2007)}]{Sudarsky07}
\bibinfo{author}{\bibfnamefont{D.}~\bibnamefont{Sudarsky}},
  \bibinfo{journal}{J. Phys. Conf. Ser.} \textbf{\bibinfo{volume}{68}},
  \bibinfo{pages}{012029} (\bibinfo{year}{2007}), \eprint{gr-qc/0612005}.

\bibitem[{\citenamefont{{de Un{\'a}nue} and {Sudarsky}}(2008)}]{US08}
\bibinfo{author}{\bibfnamefont{A.}~\bibnamefont{{de Un{\'a}nue}}}
  \bibnamefont{and}
  \bibinfo{author}{\bibfnamefont{D.}~\bibnamefont{{Sudarsky}}},
  \bibinfo{journal}{Physical Review D} \textbf{\bibinfo{volume}{78}},
  \bibinfo{pages}{043510} (\bibinfo{year}{2008}), \eprint{arXiv:0801.4702}.

\bibitem[{\citenamefont{{Le{\'o}n} and {Sudarsky}}(2010)}]{Leon10}
\bibinfo{author}{\bibfnamefont{G.}~\bibnamefont{{Le{\'o}n}}} \bibnamefont{and}
  \bibinfo{author}{\bibfnamefont{D.}~\bibnamefont{{Sudarsky}}},
  \bibinfo{journal}{Classical and Quantum Gravity}
  \textbf{\bibinfo{volume}{27}}, \bibinfo{eid}{225017} (\bibinfo{year}{2010}),
  \eprint{1003.5950}.

\bibitem[{\citenamefont{{Le{\'o}n} et~al.}(2011)\citenamefont{{Le{\'o}n}, {De
  Un{\'a}nue}, and {Sudarsky}}}]{Leon11}
\bibinfo{author}{\bibfnamefont{G.}~\bibnamefont{{Le{\'o}n}}},
  \bibinfo{author}{\bibfnamefont{A.}~\bibnamefont{{De Un{\'a}nue}}},
  \bibnamefont{and}
  \bibinfo{author}{\bibfnamefont{D.}~\bibnamefont{{Sudarsky}}},
  \bibinfo{journal}{Classical and Quantum Gravity}
  \textbf{\bibinfo{volume}{28}}, \bibinfo{pages}{155010}
  (\bibinfo{year}{2011}), \eprint{1012.2419}.

\bibitem[{\citenamefont{{Diez-Tejedor} and {Sudarsky}}(2012)}]{DT11}
\bibinfo{author}{\bibfnamefont{A.}~\bibnamefont{{Diez-Tejedor}}}
  \bibnamefont{and}
  \bibinfo{author}{\bibfnamefont{D.}~\bibnamefont{{Sudarsky}}},
  \bibinfo{journal}{JCAP} \textbf{\bibinfo{volume}{7}}, \bibinfo{eid}{045}
  (\bibinfo{year}{2012}), \eprint{1108.4928}.

\bibitem[{\citenamefont{{Landau} et~al.}(2012)\citenamefont{{Landau},
  {Sc{\'o}ccola}, and {Sudarsky}}}]{LSS12}
\bibinfo{author}{\bibfnamefont{S.~J.} \bibnamefont{{Landau}}},
  \bibinfo{author}{\bibfnamefont{C.~G.} \bibnamefont{{Sc{\'o}ccola}}},
  \bibnamefont{and}
  \bibinfo{author}{\bibfnamefont{D.}~\bibnamefont{{Sudarsky}}},
  \bibinfo{journal}{Physical Review D} \textbf{\bibinfo{volume}{85}},
  \bibinfo{eid}{123001} (\bibinfo{year}{2012}), \eprint{1112.1830}.

\bibitem[{\citenamefont{{Ca{\~n}ate} et~al.}(2013)\citenamefont{{Ca{\~n}ate},
  {Pearle}, and {Sudarsky}}}]{CPS13}
\bibinfo{author}{\bibfnamefont{P.}~\bibnamefont{{Ca{\~n}ate}}},
  \bibinfo{author}{\bibfnamefont{P.}~\bibnamefont{{Pearle}}}, \bibnamefont{and}
  \bibinfo{author}{\bibfnamefont{D.}~\bibnamefont{{Sudarsky}}},
  \bibinfo{journal}{Phys.Rev. D} \textbf{\bibinfo{volume}{87}},
  \bibinfo{eid}{104024} (\bibinfo{year}{2013}), \eprint{1211.3463}.

\bibitem[{\citenamefont{Le\'{o}n et~al.}(2014)\citenamefont{Le\'{o}n, Landau,
  and Piccirilli}}]{LLP14}
\bibinfo{author}{\bibfnamefont{G.}~\bibnamefont{Le\'{o}n}},
  \bibinfo{author}{\bibfnamefont{S.~J.} \bibnamefont{Landau}},
  \bibnamefont{and} \bibinfo{author}{\bibfnamefont{M.~P.}
  \bibnamefont{Piccirilli}}, \bibinfo{journal}{Phys. Rev.}
  \textbf{\bibinfo{volume}{D90}}, \bibinfo{pages}{083525}
  (\bibinfo{year}{2014}), \eprint{1410.1562}.

\bibitem[{\citenamefont{Le\'{o}n and Sudarsky}(2015)}]{Bispectrum}
\bibinfo{author}{\bibfnamefont{G.}~\bibnamefont{Le\'{o}n}} \bibnamefont{and}
  \bibinfo{author}{\bibfnamefont{D.}~\bibnamefont{Sudarsky}},
  \bibinfo{journal}{JCAP} \textbf{\bibinfo{volume}{1506}}, \bibinfo{pages}{020}
  (\bibinfo{year}{2015}), \eprint{1503.01417}.

\bibitem[{\citenamefont{Markkanen et~al.}(2015)\citenamefont{Markkanen,
  Rasanen, and Wahlman}}]{Markkanen2014}
\bibinfo{author}{\bibfnamefont{T.}~\bibnamefont{Markkanen}},
  \bibinfo{author}{\bibfnamefont{S.}~\bibnamefont{Rasanen}}, \bibnamefont{and}
  \bibinfo{author}{\bibfnamefont{P.}~\bibnamefont{Wahlman}},
  \bibinfo{journal}{Phys. Rev.} \textbf{\bibinfo{volume}{D91}},
  \bibinfo{pages}{084064} (\bibinfo{year}{2015}), \eprint{1407.4691}.

\bibitem[{\citenamefont{Diez-Tejedor et~al.}(2012)\citenamefont{Diez-Tejedor,
  Leon, and Sudarsky}}]{DT11a}
\bibinfo{author}{\bibfnamefont{A.}~\bibnamefont{Diez-Tejedor}},
  \bibinfo{author}{\bibfnamefont{G.}~\bibnamefont{Leon}}, \bibnamefont{and}
  \bibinfo{author}{\bibfnamefont{D.}~\bibnamefont{Sudarsky}},
  \bibinfo{journal}{Gen.Rel.Grav.} \textbf{\bibinfo{volume}{44}},
  \bibinfo{pages}{2965} (\bibinfo{year}{2012}), \eprint{1106.1176}.

\bibitem[{\citenamefont{DeWitt}(1967)}]{DeWitt67}
\bibinfo{author}{\bibfnamefont{B.~S.} \bibnamefont{DeWitt}},
  \bibinfo{journal}{Phys.Rev.} \textbf{\bibinfo{volume}{160}},
  \bibinfo{pages}{1113} (\bibinfo{year}{1967}).

\bibitem[{\citenamefont{Giulini and Kiefer}(2007)}]{Kieffer06}
\bibinfo{author}{\bibfnamefont{D.}~\bibnamefont{Giulini}} \bibnamefont{and}
  \bibinfo{author}{\bibfnamefont{C.}~\bibnamefont{Kiefer}},
  \bibinfo{journal}{Lect.Notes Phys.} \textbf{\bibinfo{volume}{721}},
  \bibinfo{pages}{131} (\bibinfo{year}{2007}), \eprint{gr-qc/0611141}.

\bibitem[{\citenamefont{Isham}(1992)}]{Isham92}
\bibinfo{author}{\bibfnamefont{C.}~\bibnamefont{Isham}},
  \emph{\bibinfo{title}{{Canonical quantum gravity and the problem of time}}}
  (\bibinfo{year}{1992}), \eprint{gr-qc/9210011}.

\bibitem[{\citenamefont{Okon and Sudarsky}(2015)}]{Elias}
\bibinfo{author}{\bibfnamefont{E.}~\bibnamefont{Okon}} \bibnamefont{and}
  \bibinfo{author}{\bibfnamefont{D.}~\bibnamefont{Sudarsky}},
  \bibinfo{journal}{Found.Phys.} \textbf{\bibinfo{volume}{45}},
  \bibinfo{pages}{461} (\bibinfo{year}{2015}), \eprint{1406.2011}.

\bibitem[{\citenamefont{Modak et~al.}(2015{\natexlab{a}})\citenamefont{Modak,
  Ort\'{i}z, Pe\~{n}a, and Sudarsky}}]{Sujoy}
\bibinfo{author}{\bibfnamefont{S.~K.} \bibnamefont{Modak}},
  \bibinfo{author}{\bibfnamefont{L.}~\bibnamefont{Ort\'{i}z}},
  \bibinfo{author}{\bibfnamefont{I.}~\bibnamefont{Pe\~{n}a}}, \bibnamefont{and}
  \bibinfo{author}{\bibfnamefont{D.}~\bibnamefont{Sudarsky}},
  \bibinfo{journal}{Gen. Rel. Grav.} \textbf{\bibinfo{volume}{47}},
  \bibinfo{pages}{120} (\bibinfo{year}{2015}{\natexlab{a}}),
  \eprint{1406.4898}.

\bibitem[{\citenamefont{Modak et~al.}(2015{\natexlab{b}})\citenamefont{Modak,
  Ort\'{i}z, Pe\~{n}a, and Sudarsky}}]{Sujoy2}
\bibinfo{author}{\bibfnamefont{S.~K.} \bibnamefont{Modak}},
  \bibinfo{author}{\bibfnamefont{L.}~\bibnamefont{Ort\'{i}z}},
  \bibinfo{author}{\bibfnamefont{I.}~\bibnamefont{Pe\~{n}a}}, \bibnamefont{and}
  \bibinfo{author}{\bibfnamefont{D.}~\bibnamefont{Sudarsky}},
  \bibinfo{journal}{Phys. Rev.} \textbf{\bibinfo{volume}{D91}},
  \bibinfo{pages}{124009} (\bibinfo{year}{2015}{\natexlab{b}}),
  \eprint{1408.3062}.

\bibitem[{\citenamefont{Jacobson and Hu}(1993)}]{jacobson}
\bibinfo{author}{\bibfnamefont{T.}~\bibnamefont{Jacobson}} \bibnamefont{and}
  \bibinfo{author}{\bibfnamefont{B.~L.} \bibnamefont{Hu}},
  \emph{\bibinfo{title}{Directions in General Reltivity}}
  (\bibinfo{publisher}{Cambridge University Press (UK)}, \bibinfo{year}{1993}).

\bibitem[{\citenamefont{Ashtekar}(2013)}]{ashtekar}
\bibinfo{author}{\bibfnamefont{A.}~\bibnamefont{Ashtekar}},
  \emph{\bibinfo{title}{{Quantum geometry and gravity: Recent advances}}}
  (\bibinfo{year}{2013}), \eprint{gr-qc/0112038}.

\bibitem[{\citenamefont{{Mariani} et~al.}(2014)\citenamefont{{Mariani},
  {Bengochea}, and {Leon}}}]{MBL14}
\bibinfo{author}{\bibfnamefont{M.}~\bibnamefont{{Mariani}}},
  \bibinfo{author}{\bibfnamefont{G.~R.} \bibnamefont{{Bengochea}}},
  \bibnamefont{and} \bibinfo{author}{\bibfnamefont{G.}~\bibnamefont{{Leon}}},
  \emph{\bibinfo{title}{{Inflationary gravitational waves in collapse scheme
  models}}} (\bibinfo{year}{2014}), \eprint{1412.6471}.

\bibitem[{\citenamefont{Leon and Bengochea}(2015)}]{LB15}
\bibinfo{author}{\bibfnamefont{G.}~\bibnamefont{Leon}} \bibnamefont{and}
  \bibinfo{author}{\bibfnamefont{G.~R.} \bibnamefont{Bengochea}},
  \emph{\bibinfo{title}{{Emergence of inflationary perturbations in the CSL
  model}}} (\bibinfo{year}{2015}), \eprint{1502.04907}.

\bibitem[{\citenamefont{Bassi and Ghirardi}(2003)}]{bassi2003}
\bibinfo{author}{\bibfnamefont{A.}~\bibnamefont{Bassi}} \bibnamefont{and}
  \bibinfo{author}{\bibfnamefont{G.~C.} \bibnamefont{Ghirardi}},
  \bibinfo{journal}{Phys.Rept.} \textbf{\bibinfo{volume}{379}},
  \bibinfo{pages}{257} (\bibinfo{year}{2003}), \eprint{quant-ph/0302164}.

\bibitem[{\citenamefont{Penrose}(1996)}]{penrose1996}
\bibinfo{author}{\bibfnamefont{R.}~\bibnamefont{Penrose}},
  \bibinfo{journal}{Gen.Rel.Grav.} \textbf{\bibinfo{volume}{28}},
  \bibinfo{pages}{581} (\bibinfo{year}{1996}).

\bibitem[{\citenamefont{Diosi}(1987)}]{diosi1987}
\bibinfo{author}{\bibfnamefont{L.}~\bibnamefont{Diosi}},
  \bibinfo{journal}{Phys.Lett.} \textbf{\bibinfo{volume}{A120}},
  \bibinfo{pages}{377} (\bibinfo{year}{1987}).

\bibitem[{\citenamefont{Mollerach and Matarrese}(1997)}]{Mollerach1997}
\bibinfo{author}{\bibfnamefont{S.}~\bibnamefont{Mollerach}} \bibnamefont{and}
  \bibinfo{author}{\bibfnamefont{S.}~\bibnamefont{Matarrese}},
  \bibinfo{journal}{Phys.Rev.} \textbf{\bibinfo{volume}{D56}},
  \bibinfo{pages}{4494} (\bibinfo{year}{1997}), \eprint{astro-ph/9702234}.

\bibitem[{\citenamefont{Mollerach}(1998)}]{Mollerach1998}
\bibinfo{author}{\bibfnamefont{S.}~\bibnamefont{Mollerach}},
  \bibinfo{journal}{Phys.Rev.} \textbf{\bibinfo{volume}{D57}},
  \bibinfo{pages}{1303} (\bibinfo{year}{1998}), \eprint{astro-ph/9708196}.

\bibitem[{\citenamefont{Mollerach et~al.}(2004)\citenamefont{Mollerach, Harari,
  and Matarrese}}]{Mollerach2003}
\bibinfo{author}{\bibfnamefont{S.}~\bibnamefont{Mollerach}},
  \bibinfo{author}{\bibfnamefont{D.}~\bibnamefont{Harari}}, \bibnamefont{and}
  \bibinfo{author}{\bibfnamefont{S.}~\bibnamefont{Matarrese}},
  \bibinfo{journal}{Phys.Rev.} \textbf{\bibinfo{volume}{D69}},
  \bibinfo{pages}{063002} (\bibinfo{year}{2004}), \eprint{astro-ph/0310711}.

\bibitem[{\citenamefont{Osano et~al.}(2007)\citenamefont{Osano, Pitrou, Dunsby,
  Uzan, and Clarkson}}]{Osano2006}
\bibinfo{author}{\bibfnamefont{B.}~\bibnamefont{Osano}},
  \bibinfo{author}{\bibfnamefont{C.}~\bibnamefont{Pitrou}},
  \bibinfo{author}{\bibfnamefont{P.}~\bibnamefont{Dunsby}},
  \bibinfo{author}{\bibfnamefont{J.-P.} \bibnamefont{Uzan}}, \bibnamefont{and}
  \bibinfo{author}{\bibfnamefont{C.}~\bibnamefont{Clarkson}},
  \bibinfo{journal}{JCAP} \textbf{\bibinfo{volume}{0704}}, \bibinfo{pages}{003}
  (\bibinfo{year}{2007}), \eprint{gr-qc/0612108}.

\bibitem[{\citenamefont{Ananda et~al.}(2007)\citenamefont{Ananda, Clarkson, and
  Wands}}]{Ananda2006}
\bibinfo{author}{\bibfnamefont{K.~N.} \bibnamefont{Ananda}},
  \bibinfo{author}{\bibfnamefont{C.}~\bibnamefont{Clarkson}}, \bibnamefont{and}
  \bibinfo{author}{\bibfnamefont{D.}~\bibnamefont{Wands}},
  \bibinfo{journal}{Phys.Rev.} \textbf{\bibinfo{volume}{D75}},
  \bibinfo{pages}{123518} (\bibinfo{year}{2007}), \eprint{gr-qc/0612013}.

\bibitem[{\citenamefont{Baumann et~al.}(2007)\citenamefont{Baumann, Steinhardt,
  Takahashi, and Ichiki}}]{Baumann2007}
\bibinfo{author}{\bibfnamefont{D.}~\bibnamefont{Baumann}},
  \bibinfo{author}{\bibfnamefont{P.~J.} \bibnamefont{Steinhardt}},
  \bibinfo{author}{\bibfnamefont{K.}~\bibnamefont{Takahashi}},
  \bibnamefont{and} \bibinfo{author}{\bibfnamefont{K.}~\bibnamefont{Ichiki}},
  \bibinfo{journal}{Phys.Rev.} \textbf{\bibinfo{volume}{D76}},
  \bibinfo{pages}{084019} (\bibinfo{year}{2007}), \eprint{hep-th/0703290}.

\bibitem[{\citenamefont{Le\'{o}n et~al.}(2015)\citenamefont{Le\'{o}n, Landau,
  and Piccirilli}}]{Pia2015}
\bibinfo{author}{\bibfnamefont{G.}~\bibnamefont{Le\'{o}n}},
  \bibinfo{author}{\bibfnamefont{S.~J.} \bibnamefont{Landau}},
  \bibnamefont{and} \bibinfo{author}{\bibfnamefont{M.~P.}
  \bibnamefont{Piccirilli}}, \bibinfo{journal}{Eur. Phys. J.}
  \textbf{\bibinfo{volume}{C75}}, \bibinfo{pages}{393} (\bibinfo{year}{2015}),
  \eprint{1502.00921}.

\bibitem[{\citenamefont{Acquaviva et~al.}(2003)\citenamefont{Acquaviva,
  Bartolo, Matarrese, and Riotto}}]{Acquaviva2002}
\bibinfo{author}{\bibfnamefont{V.}~\bibnamefont{Acquaviva}},
  \bibinfo{author}{\bibfnamefont{N.}~\bibnamefont{Bartolo}},
  \bibinfo{author}{\bibfnamefont{S.}~\bibnamefont{Matarrese}},
  \bibnamefont{and} \bibinfo{author}{\bibfnamefont{A.}~\bibnamefont{Riotto}},
  \bibinfo{journal}{Nucl.Phys.} \textbf{\bibinfo{volume}{B667}},
  \bibinfo{pages}{119} (\bibinfo{year}{2003}), \eprint{astro-ph/0209156}.

\bibitem[{\citenamefont{{Planck Collaboration}
  et~al.}(2015)\citenamefont{{Planck Collaboration}, {Ade}, {Aghanim},
  {Arnaud}, {Arroja}, {Ashdown}, {Aumont}, {Baccigalupi}, {Ballardini},
  {Banday} et~al.}}]{Planckinflation15}
\bibinfo{author}{\bibnamefont{{Planck Collaboration}}},
  \bibinfo{author}{\bibfnamefont{P.~A.~R.} \bibnamefont{{Ade}}},
  \bibinfo{author}{\bibfnamefont{N.}~\bibnamefont{{Aghanim}}},
  \bibinfo{author}{\bibfnamefont{M.}~\bibnamefont{{Arnaud}}},
  \bibinfo{author}{\bibfnamefont{F.}~\bibnamefont{{Arroja}}},
  \bibinfo{author}{\bibfnamefont{M.}~\bibnamefont{{Ashdown}}},
  \bibinfo{author}{\bibfnamefont{J.}~\bibnamefont{{Aumont}}},
  \bibinfo{author}{\bibfnamefont{C.}~\bibnamefont{{Baccigalupi}}},
  \bibinfo{author}{\bibfnamefont{M.}~\bibnamefont{{Ballardini}}},
  \bibinfo{author}{\bibfnamefont{A.~J.} \bibnamefont{{Banday}}},
  \bibnamefont{et~al.}, \bibinfo{journal}{ArXiv e-prints}
  (\bibinfo{year}{2015}), \eprint{1502.02114}.

\bibitem[{\citenamefont{Starobinsky}(1980)}]{starobinsky}
\bibinfo{author}{\bibfnamefont{A.~A.} \bibnamefont{Starobinsky}},
  \bibinfo{journal}{Phys.Lett.} \textbf{\bibinfo{volume}{B91}},
  \bibinfo{pages}{99} (\bibinfo{year}{1980}).

\bibitem[{\citenamefont{Mukhanov and Chibisov}(1981)}]{mukhanov81}
\bibinfo{author}{\bibfnamefont{V.~F.} \bibnamefont{Mukhanov}} \bibnamefont{and}
  \bibinfo{author}{\bibfnamefont{G.~V.} \bibnamefont{Chibisov}},
  \bibinfo{journal}{JETP Lett.} \textbf{\bibinfo{volume}{33}},
  \bibinfo{pages}{532} (\bibinfo{year}{1981}).

\bibitem[{\citenamefont{Starobinsky}(1983)}]{starobinsky2}
\bibinfo{author}{\bibfnamefont{A.~A.} \bibnamefont{Starobinsky}},
  \bibinfo{journal}{Sov.Astron.Lett.} \textbf{\bibinfo{volume}{9}},
  \bibinfo{pages}{302} (\bibinfo{year}{1983}).

\bibitem[{\citenamefont{Kehagias et~al.}(2014)\citenamefont{Kehagias, Dizgah,
  and Riotto}}]{Kehagias2013}
\bibinfo{author}{\bibfnamefont{A.}~\bibnamefont{Kehagias}},
  \bibinfo{author}{\bibfnamefont{A.~M.} \bibnamefont{Dizgah}},
  \bibnamefont{and} \bibinfo{author}{\bibfnamefont{A.}~\bibnamefont{Riotto}},
  \bibinfo{journal}{Phys.Rev.} \textbf{\bibinfo{volume}{D89}},
  \bibinfo{pages}{043527} (\bibinfo{year}{2014}), \eprint{1312.1155}.

\end{thebibliography}
\bibliographystyle{apsrev}

\end{document}